\documentclass[journal=jctcce,manuscript=article]{achemso}
\usepackage{amsmath}
\usepackage{nameref}
\usepackage{blindtext}
\usepackage[dvipsnames]{xcolor}
\usepackage[utf8]{inputenc}
\usepackage{graphicx} 
\usepackage{bm}	
\usepackage{color}                     
\usepackage{xcolor}
\usepackage{epsfig}
\usepackage{amsmath} 
\usepackage{amssymb} 
\usepackage{longtable} 
\usepackage{float}
\usepackage{mathtools}
\usepackage{xparse}
\usepackage{hyperref}
\usepackage{times}
\usepackage[normalem]{ulem}
\usepackage{sidecap}
\usepackage{subcaption}
\usepackage{appendix} 
\usepackage{multirow}
\sidecaptionvpos{figure}{t}
\usepackage{comment}
\usepackage{tikz,forest}
\usetikzlibrary{arrows.meta}
\usepackage[ruled,vlined]{algorithm2e}
\usepackage[version=4]{mhchem}

\SetKwBlock{myInit}{Initialization}{end}
\SetKwBlock{myCDGW}{CD-$GW$}{end}

\setkeys{acs}{etalmode=truncate,maxauthors=10}
\setkeys{acs}{articletitle=true}

\title{Scalable Molecular GW Calculations: Valence and Core Spectra}

\author{Daniel Mejia-Rodriguez}
\email{daniel.mejia@pnnl.gov}
\affiliation{Environmental Molecular Sciences Laboratory, Pacific Northwest National Laboratory, Richland, WA 99352, USA}

\author{Alexander Kunitsa}
\email{aakunitsa@gmail.com}
\affiliation{Department of Chemistry, University of Illinois at Urbana-Champaign, 600 S. Mathews Avenue, Urbana, Illinois, 61801, USA}
\altaffiliation{Present Address: Zapata Computing, Inc., 100 Federal Street,
Boston, MA 02110, USA}

\author{Edoardo Aprà}
\email{edoardo.apra@pnnl.gov}
\affiliation{Environmental Molecular Sciences Laboratory, Pacific Northwest National Laboratory, Richland, WA 99352, USA}

\author{Niranjan Govind}
\email{niri.govind@pnnl.gov}
\affiliation{Physical and Computational Sciences Directorate, Pacific Northwest National Laboratory, Richland, WA 99352, USA}

\newcommand{\eri}[2]{\ensuremath{\left( #1 \big\vert #2 \right) }}
\newcommand{\onlinecite}[1]{\nocite{#1}\citenum{#1}}

\begin{document}
\maketitle
\newpage
\begin{abstract}
We present a scalable implementation of the $GW$ approximation
using Gaussian atomic orbitals to study the valence and core ionization spectroscopies of molecules. The implementation of the standard spectral
decomposition approach to the screened Coulomb interaction, 
as well as a contour deformation method are described. We have implemented both of these approaches using the robust variational fitting approximation to the four-center electron repulsion integrals. We have utilized the MINRES solver with the contour deformation approach to reduce the computational scaling by one order of magnitude. A complex heuristic in the quasiparticle equation solver further allows a speed-up of the computation of core and semi-core ionization energies. Benchmark tests using the GW100 and CORE65 datasets and the carbon 1{\it s} binding energy of the well-studied ethyl trifluoroacetate, or ESCA molecule, were performed to validate the accuracy of our implementation. We also demonstrate and discuss the parallel performance and computational scaling of our implementation using a range of water clusters of increasing size.


\end{abstract}

\section{Introduction}
\label{intr}
Many-body perturbation theory (MBPT) has been extensively demonstrated\cite{Szabo:1996,Fetter:2003,Shavitt:2009} over the years as a worthy alternative to density functional theory (DFT). This has resulted in a number of useful approximations that are widely used in solid-state physics and quantum chemistry.
Specifically, within the framework of MBPT, the $GW$ approximation to the self-energy $\Sigma$\cite{Hedin:1965:A796},
has been used with considerable success over the last three decades and has widespread reputation as an accurate and efficient method for the prediction of band-structures in solids. In recent years there has been growing interest in the $GW$ method applied to molecular and finite systems (see reviews \citenum{Aryasetiawan:1998:237,Onida:2002:601,Leng:2016:532,Cances:2016:1650008,Reining:2018:e1344,Golze:2019:377} and references therein). 

The $GW$ quasi-particle energies, unlike the Kohn-Sham (KS) single-particle energies, can be used to calculate properties associated with charged excitations (i.e. electron addition and removal) that can be compared to photoemission and inverse-photoemission spectroscopy.  Other key features include the absence of empirical parameters, \emph{ab initio} inclusion of dynamical electron correlation, appropriate long range behavior of the electron-hole interaction and a well-defined physical meaning of the quasi-particle energies. 
In addition, the moderate computational cost of the $GW$ approach, between DFT and traditional quantum chemistry correlated approaches, makes it an attractive first-principles theory for charged excitation energies. 
The $GW$ quasi-particle energies also serve as a starting point to describe neutral transitions (for example, optical and x-ray absorption spectroscopies) via the Bethe-Salpeter (BSE) formalism \cite{Salpeter:1951:1232}, which introduces electron-hole interaction effects at the same level of theory.

The simplest and most popular form of the $GW$ approximation is the first-order $G_0W_0$ approach, which is performed as a post-processing or one-shot step to a KS (local, semilocal, hybrid or pure Hartree-Fock (HF)) reference calculation. Here, the non-interacting Green’s function $G$ is diagonal in the KS (or HF) eigenstate basis, and only the diagonal matrix elements of the self-energy are needed to evaluate the quasi-particle energy corrections to first order.
Some of the issues (for example, the KS or HF reference dependence) of the one-shot approach can be improved with different levels
of refinement, such as the eigenvalue (ev) self-consistent approaches where the quasi-particle eigenvalues are used to iteratively update the Green's Function $G$ (ev$GW_0$) or both $G$ and $W$ (ev$GW$) or the more complex self-consistent (sc) approaches, where both the orbitals and quasi-particle eigenvalues are calculated and iterated to self-consistency in $G$ (sc$GW_0$) or both $G$ and $W$ (sc$GW$), respectively.\cite{Faleev:2004:126406,Stan:2009:114105,Caruso:2012:081102}
Consequently, the $GW$ approximation is now an integral part of electronic structure theory and is available in many widely used electronic structure codes like Turbomole \cite{VanSetten2013,Kaplan:2016:2528,Holzer:2019:204116,TURBOMOLE}, FHI-Aims \cite{Caruso:2013:075105} , CP2K \cite{Wilhelm:2018:306,CP2K}, ADF \cite{Forster:2020:7381}, molGW \cite{Bruneval2016}, BerkeleyGW\cite{BerkeleyGW}, \textsc{Quantum} ESPRESSO\cite{QE:2009,QE:2017,GWL:2009,GWL:2010}, \textsc{Abinit}\cite{ABINIT:2020}, VASP\cite{VASP:1993,VASP:1994,VASP:1996,VASP:1996:11169}, Yambo \cite{Yambo:2009,Yambo:2019}, WEST\cite{WEST}, Elk \cite{Elk}, PySCF \cite{pyscf:2018,pyscf:2020}, GPAW\cite{GPAW:2005,GPAW:2010,Hueser:2013:235132}, WIEN2k \cite{WIEN2k,GAP,GAP2}, Questaal \cite{Questaal}, Spex \cite{Spex}.

In this paper, we describe 
our implementation of the $GW$ approximation based on the software infrastructure of the
open-source \textsc{NWChem} computational chemistry program \cite{NWChem} within the Gaussian basis set framework for molecular and finite systems. Our implementation can use either the spectral decomposition (SD) or the contour-deformation (CD) methods. Both approaches are suitable to describe valence and core ionization spectra. The rest of the paper is organized as follows: For completeness, in Section \ref{theory}, we describe the theoretical framework of the $GW$ approximation and the theoretical background of both of the approaches we have implemented to obtain the
screened Coulomb interaction. Section \ref{impl} describes
the details of the implementation, with special emphasis on the
contour deformation approach. Section \ref{res} shows
benchmark results for both core and valence ionizations
by comparing with the GW100 \cite{vanSetten:2015:5665} and CORE65\cite{Golze:2020:1840} datasets as well as computations of the carbon 1{\it s} binding energies of the ESCA molecule. Parallel
scalability is presented and discussed in Section \ref{par-perf}. Finally, a brief
summary and outlook is presented in Section \ref{summ}.

\section{Theory}
\label{theory}

\subsection{Overview of the Hedin Equations and Derivation of the \texorpdfstring{$GW$}{GW} Approximation}
The central object of the $GW$ theory is the one-particle Green's  function $G$ describing particle and hole scattering in the interacting many-body system. Formally $G$ is defined in terms of time-ordered products of creation ($\hat{\psi}^{\dagger}$) and annihilation ($\hat{\psi}$) operators in the Heisenberg representation~\cite{Fetter:2003}
\begin{align*}
    iG(1,2) &= \langle \Psi_0 | T[\hat{\psi}_H(1)\hat{\psi}^{\dagger}_H(2)] | \Psi_0 \rangle\\
    &= \theta(t_1 - t_2)\langle \Psi_0 | \hat{\psi}_H(1)\hat{\psi}^{\dagger}_H(2) | \Psi_0 \rangle -  \theta(t_2 - t_1) \langle \Psi_0 | \hat{\psi}^{\dagger}_H(2) \hat{\psi}_H(1)| \Psi_0 \rangle
\end{align*}
where $| \Psi_0 \rangle$ is an exact ground state satisfying the time-independent Schr\"odinger equation $H| \Psi_0 \rangle = E_0 | \Psi_0 \rangle$, $i=1,2,...$ refers to a combined space-time coordinate $(r_i, t_i)$ and
$\theta$ is the Heaviside step function under the half maximum convention.  
For simplicity, spin variables have been omitted in the above definition. 
The $GW$ approximation can be derived from the Hedin's equations\cite{Hedin:1965:A796}
by substituting the vertex function $\Gamma$ with the product of the two delta functions
$\Gamma(1,2,3) = \delta(1-2) \delta(2-3)$ which amounts to the first order approximation to the self-energy $\Sigma$ in terms of the screened Coulomb potential $W$:
\begin{align}
\Sigma(1,2) &= i G(1, 2) W(1, 2^{+}), \label{eqn:gw_eqns_se}
\end{align}
where ``+" indicates chronological ordering (e.g. $t_1^{+} = t_1 + \eta$, $\eta \to 0$). From a physical standpoint $W$ describes the interaction between dynamically screened quasi-particles as reflected in its definition in terms of the inverse dielectric function $\epsilon^{-1}(1,2)$ and bare Coulomb potential $v(1,2) = \delta(t_1 - t_2) \frac{1}{|r_1 - r_2|}$ (assuming instantaneous and spin-independent interaction in the non-relativistic case) 
\begin{align}
W(1, 2) = \int d3 \epsilon^{-1}(1,3)v(3,2).
\end{align}
It can be shown that $W$ satisfies a Dyson-like equation
\begin{align}
W(1,2) &= v(1,2) + \int d(34) v(1,3) P(3,4) W(4,2),  \label{eqn:gw_eqns_w}    
\end{align}
connecting it to the irreducible polarizability $P$ describing the density response with respect to the total potential (accounting for both induced and external contributions):
\begin{align}
P(1,2) &= -i G(1,2) G(2,1^{+})\label{eqn:gw_eqns_pol}
\end{align}
Along with the Dyson equation for the interacting Green's function (in terms of the Hartree Green's function, $G_H$)
\begin{align}
G(1,2) &= G_H(1,2) + \int d(34) G_H(1,3) \Sigma(3,4) G(4,2) \label{eqn:gw_eqns_dyson}    
\end{align}
equations~\ref{eqn:gw_eqns_se},~\ref{eqn:gw_eqns_pol}, and~\ref{eqn:gw_eqns_w} form the basis of $GW$ approach. The crux of $GW$ is the evaluation of the screened Coulomb interaction $W(1,2)$ which can, in principle, be obtained by solving Eq.~\ref{eqn:gw_eqns_w}.
In practical applications it is more efficient to express $W$ in terms of the  reducible polarizability, or density response function, $\chi$ describing the perturbation of the electronic density by a time- dependent external potential:
\begin{align}
    \chi(1,2) = P(1,2) + \int d(34)P(1,3)v(3,4)\chi(4,2)\label{eqn:chi}
\end{align}
Combining the above equation with~\ref{eqn:gw_eqns_w} one effectively obtains a closed form expression for $W$ in terms of $\chi$ the high quality approximations which are available in standard electronic structure packages:
\begin{align}
\label{eqn:w_density}
W(1,2) &= v(1,2) + \int d(34) v(1,3) \chi(3,4) v(4,2)    
\end{align}
In the non-relativistic case Eq.~\ref{eqn:w_density} can be further simplified:
\begin{align}
    W(1,2) &= v(r_1,r_2)\delta(t_1 - t_2)  + \int d(r_3r_4) v(r_1,r_3) \chi(r_3,r_4; t_1 - t_2) v(r_4,r_2)
\end{align}
In the frequency domain the equation for the self-energy is transformed as follows:
\begin{align}
\Sigma(r_1,r_2, \omega) = \frac{i}{2\pi} \int_{-\infty}^{+\infty} e^{i\xi \eta} G(r_1, r_2, \omega + \xi) W(r_1, r_2, \xi) d\xi,   
\end{align}
under the usual assumption of $\eta \to 0^+$. Combining the previous equation with the expression for screened Coulomb interaction one obtains:

\begin{align}
 \Sigma(r_1,r_2, \omega) = \frac{i}{2\pi} v(r_1, r_2) \int_{-\infty}^{+\infty} e^{i\xi \eta} G(r_1, r_2, \omega + \xi) d\xi \ + \nonumber \\  \frac{i}{2\pi} \int_{-\infty}^{+\infty} d \xi \int d(r_3r_4) e^{i\xi \eta} G(r_1, r_2, \omega + \xi) v(r_1,r_3) \chi(r_3,r_4; \xi) v(r_4,r_2)  
 \label{eq:sigma}
\end{align}
where the first contribution can be readily recognized as an \textit{exchange} part of the self-energy since $-\rho(r_1, r_2) = \frac{i}{2\pi} \int_{-\infty}^{+\infty} e^{i\xi \eta} G(r_1, r_2, \xi) d\xi$ is the one particle density matrix. 

\subsection{Spectral Decomposition}
One way of obtaining $\Sigma(\omega)$ is via the spectral
decomposition (SD) of the density response function $\chi$ in the 
random phase approximation (RPA):

\begin{equation}
    \chi(r_1,r_2,\omega) = \sum\limits_s n_s(r_1) n_s(r_2) \left( \frac{1}{\omega - \Omega_s + i\eta} - \frac{1}{\omega + \Omega_s - i\eta} \right)
\end{equation}
where $n_s$ are transition densities and $\Omega_s$ are the
charge-neutral excitations. These can be obtained by solving the
Casida equations:

\begin{align}
\label{eqn:rpa}
\begin{bmatrix}A & B\\-B & -A\end{bmatrix} \begin{bmatrix}X_s\\Y_s\end{bmatrix} = \Omega_s \begin{bmatrix}X_s\\Y_s\end{bmatrix},
\end{align}
where, for closed-shells, $A_{ia, jb} = \delta_{ij} \delta_{ab} (\epsilon_a - \epsilon_i) + 2(ia|jb)$, $B_{ia, jb} = 2(ia|bj)$. The open-shell equations can be written by explicitly exposing the spin blocks as
\begin{align}
\label{eqn:rpaos}
\begin{bmatrix}A_{\uparrow\uparrow} & B_{\uparrow\uparrow} & B_{\uparrow\downarrow} & B_{\uparrow\downarrow} \\-B_{\uparrow\uparrow} & -A_{\uparrow\uparrow} & -B_{\uparrow\downarrow} & -B_{\uparrow\downarrow} \\ B_{\downarrow\uparrow} & B_{\downarrow\uparrow} & A_{\downarrow\downarrow} & B_{\downarrow\downarrow} \\ -B_{\downarrow\uparrow} & -B_{\downarrow\uparrow} & -B_{\downarrow\downarrow} & -A_{\downarrow\downarrow}  \end{bmatrix} \begin{bmatrix}X_{\uparrow}^s\\Y_{\uparrow}^s \\ X_{\downarrow}^s\\ Y_{\downarrow}^s \end{bmatrix} =  \Omega_s\begin{bmatrix} X_{\uparrow}^s\\ Y_{\uparrow}^s \\ X_{\downarrow}^s \\ Y_{\downarrow}^s \end{bmatrix},
\end{align}
and where $A_{ia\sigma,jb\sigma} = \delta_{ij}\delta_{ab}(\epsilon_{a\sigma}-\epsilon_{i\sigma}) + \eri{ia\sigma}{jb\sigma}$ and $B_{ia\sigma_1,jb\sigma_2} = \eri{ia\sigma_1}{jb\sigma_2}$.

The matrix elements of $W$ in the orbital product basis are then expressed in terms of the full set of eigenvectors $(X_s, Y_s)$ and corresponding neutral excitation energies $\Omega_s$:
\begin{align}
\label{eqn:analytic_w}
  W_{mn\sigma_1, op\sigma_2} (\omega) = (mn\sigma_1|op\sigma_2) + \sum_s \omega_{mn\sigma_1}^{s}\omega_{op\sigma_2}^{s} \times \big( \frac{1}{\omega - \Omega_s + i\eta} -  \frac{1}{\omega + \Omega_s - i\eta} \big), 
\end{align}
where $\omega_{mn\sigma}^{s} = \sum_{ia\sigma}(mn\sigma_1|ia\sigma)(X_{ia\sigma}^{s} + Y_{ia\sigma}^{s})$. Eq.~\ref{eqn:analytic_w} is a key component of the spectral decomposition approach and often serves as a starting point for deriving approximate $GW$ and BSE methods. In particular, the matrix elements of the self-energy can be directly obtained by contracting $ W_{mn\sigma_1, op\sigma_2} (\omega)$ with the KS Green's function:
\begin{align}
\label{eqn:ks_gf}
G_{\sigma\sigma\prime}^{KS}(r_1, r_2, \omega) = \delta_{\sigma\sigma\prime} \sum_k \frac{\phi_{k\sigma}(r_1)\phi_{k\sigma}^{*}(r_2) }{\omega - \epsilon_{k\sigma} + i\eta \times sign(\epsilon_{k\sigma} - \mu)}    
\end{align}
with $\mu$ being the Fermi-level of the system.

Integration is performed by closing the contour in the upper half of the complex plane such that

\begin{align*}
\int_{-\infty}^{+\infty} \frac{d\xi e^{i\xi \eta}}{(\xi + \omega - \epsilon_{k\sigma} + i\eta \times sign(\epsilon_{k\sigma} - \mu))(\xi - \Omega_s + i\eta)} = \begin{cases} 0 &\mbox{if } \epsilon_{k\sigma} > \mu\\
-\frac{2\pi i}{\omega - \epsilon_{k\sigma} + \Omega_s - 2i\eta} &\mbox{otherwise }\end{cases}\\
\int_{-\infty}^{+\infty} \frac{d\xi e^{i\xi \eta}}{(\xi + \omega - \epsilon_{k\sigma} + i\eta \times sign(\epsilon_{k\sigma} - \mu))(\xi + \Omega_s - i\eta)} = \begin{cases} \frac{2\pi i}{\omega - \epsilon_{k\sigma} - \Omega_s + 2i\eta} &\mbox{if } \epsilon_{k\sigma} > \mu\\
0 &\mbox{otherwise }\end{cases}
\end{align*}
The final expression for the self-energy in terms of tensor contractions is presented below:

\begin{align}
\Sigma_{nn\sigma}(\omega) = \frac{i}{2\pi} \int_{-\infty}^{+\infty} d \xi e^{i\xi \eta} \sum\limits_i \frac{W_{ni\sigma, in\sigma}(\xi)}{\xi + \omega - \epsilon_{i\sigma} - i\eta} + \sum\limits_a \frac{W_{na\sigma, an\sigma}(\xi)}{\xi + \omega - \epsilon_{a\sigma} + i\eta}\\
=-\sum\limits_i (ni\sigma|in\sigma) + \sum\limits_{is} \frac{\omega_{ni\sigma}^s\omega_{in\sigma}^s}{\omega - \epsilon_{i\sigma} + \Omega_s - 2i\eta} + \sum\limits_{as} \frac{\omega_{na\sigma}^s\omega_{an\sigma}^s}{\omega - \epsilon_{a\sigma} - \Omega_s + 2i\eta}    
\end{align}

The most expensive part of the spectral decomposition approach is the
computation of the neutral excitations $\Omega_s$ and the associated
eigenvectors $X_s$ and $Y_s$, which formally scales as $\mathcal{O}(N^6)$.
One way of avoiding this steep computational cost is by means of the
contour-deformation technique described below.

\subsection{Contour Deformation}
Instead of using the density response function $\chi$ in order to obtain
the screened Coulomb interaction, the contour deformation (CD) approach uses
the independent-particle irreducible polarizability $\chi_0$, which has a simple
sum-over-states representation \cite{Adler:1962,Wiser:1963}.
However, the integration in Eq. \ref{eq:sigma} can no longer be
solved analytically. Following Ref. \onlinecite{Golze:2018:4856}, the self-energy is decomposed as
\begin{equation}
    \Sigma(r_1,r_2,\omega) = R(r_1,r_2,\omega) - I(r_1,r_2,\omega)
\end{equation}
where the contour integral $R(r_1,r_2,\omega)$ and the integral
over the imaginary axis $I(r_1,r_2,\omega)$ are defined as
\begin{equation}
    R(r_1,r_2,\omega) := \frac{i}{2\pi} \oint d\xi \: G(r_1,r_2,\omega+\xi)W(r_1,r_2,\xi)
\end{equation}
\begin{equation}
    I(r_1,r_2,\omega) := \frac{1}{2\pi} \int\limits_{-\infty}^{\infty} d\xi \: G(r_1,r_2,\omega+i\xi)W(r_1,r_2,i\xi) \label{eqn:I(w)}
\end{equation}
The contour integral is evaluated using the residue theorem
by choosing the contours in such a way that only the poles of
$G_0$ are enclosed, yielding
\begin{eqnarray}
    R(r_1,r_2,\omega) = &-& \sum\limits_i \phi_i(r_1) \phi_i(r_2) W(r_1,r_2,\epsilon_i - \omega + i\eta)\theta(\epsilon_i -\omega) \nonumber \\
    &+& \sum\limits_a \phi_a(r_1) \phi_a(r_2) W(r_1,r_2,\epsilon_a - \omega - i\eta)\theta(\omega - \epsilon_a)
    \label{eq:R}
\end{eqnarray}
%

The integral over the imaginary axis $I(r_1,r_2,\omega)$ is obtained 
by inserting Eq.~\ref{eqn:ks_gf} into
Eq.~\ref{eqn:I(w)},
\begin{equation}
  I(r_1,r_2,\omega) = \frac{1}{2\pi}\sum\limits_m   \int\limits_{-\infty}^{\infty} d\xi \frac{\phi_m(r_1)\phi_m(r_2) W(r_1,r_2,i\xi)}{\omega + i\xi  - \epsilon_{m} + sign(\epsilon_k - \mu)} \; ,
  \label{eq:I}
\end{equation}

The diagonal matrix elements for the self-energy in
the MO basis are then obtained as
\begin{eqnarray}
    \Sigma_{nn\sigma}(\omega) &=& -\sum\limits_i W_{ni\sigma,ni\sigma}(\epsilon_{i\sigma}-\omega+i\eta)\theta(\epsilon_{i\sigma}-\omega) \nonumber \\
    & & + \sum\limits_a W_{na\sigma,na\sigma}(\epsilon_{a\sigma}-\omega-i\eta)\theta(\omega-\epsilon_{a\sigma}) \nonumber \\
    & & -\frac{1}{2\pi} \sum\limits_m \int\limits_{-\infty}^{\infty} d\xi \frac{W_{nm\sigma,nm\sigma}(i\xi)}{i\xi  - \epsilon_{m\sigma} + sign(\epsilon_k - \mu)} \\
    &=& R_{nn\sigma}(\omega) - I_{nn\sigma}(\omega)
\end{eqnarray}

\section{Implementation}
\label{impl}
\subsection{Variational Fitting Approximation}

We have implemented the $GW$ and $evGW$ methods
using the 
robust variational fitting (RVF) technique
\cite{Whitten:1973:4496,Dunlap:1979:3396,Dunlap:2000:37}.
Within RVF, the two-particle four-center electron repulsion integral (ERI)
\begin{equation}
    \eri{ai}{bj} := \iint d(r_1r_2) \phi_a(r_1)\phi_i(r_1) v(r_1,r_2) \phi_b(r_2) \phi_j(r_2)
\end{equation}
is approximated as
\begin{equation}
    \eri{ai}{bj} \approx \eri{\widetilde{ai}}{bj} + \eri{ai}{\widetilde{bj}} - \eri{\widetilde{ai}}{\widetilde{bj}} \label{eqn:rvf}
\end{equation}
with 
\begin{equation}
    \widetilde{\phi_i(r)\phi_a(r)} := \sum\limits_{P} C_{ia}^{P} f_P(r) 
\end{equation}
and $f_P(r)$ an atom-centered auxiliary function. The fitting coefficients
$C_{ia}^P$ are obtained by minimizing the squared norm, $\tau_{ia}^2$,
of the residual
\begin{equation}
    \Delta_{ia}(r) := \phi_i(r)\phi_a(r) - \widetilde{\phi_i(r)\phi_a(r)}
\end{equation}
in a given metric $\Omega(r_1,r_2)$. The minimization can be further
constrained to preserve the charge carried by the orbital
product $\phi_i(r)\phi_a(r)$. 

The choice of the metric influences 
the size of the auxiliary basis set needed to achieve certain accuracy,
the number of non-negligible ERIs, and the speed in which the
ERIs are obtained. Typical metrics include Coulomb, $v(r_1,r_2)$, 
short-ranged Coulomb, $v(r_1,r_2)\mathrm{erfc}(\mu r_{12})$,
truncated Coulomb $v(r_1,r_2)\theta(r_C - r_{12})$, and overlap,
$\delta(r_1,r_2)$. The Coulomb metric yields the highest accuracy
achievable with a given auxiliary basis set at the expense
of more non-negligible ERIs. In contrast, the overlap metric
is the least accurate but sparsest one.

Note that Eq.~\ref{eqn:rvf} reduces to the standard
resolution-of-the-identity (RI) \cite{Vahtras:1993:514} formula
only when the fitting coefficients $C_{ia}^P$ are obtained via
an unconstrained fit. This distinction is
crucial since, in general, the error in approximating $\eri{ia}{jb}$ 
introduced by the RI approximation is linear in $\tau_{ia}$
and $\tau_{jb}$, while that of RVF is bilinear
$\tau_{ia}\tau_{jb}$ \cite{Dunlap:2000:37,Wirz:2017:4897}.

Local fitting procedures, either using a local metric or restricting
the centers which contribute auxiliary functions to describe a
given orbital pair, have been recently used in $GW$ implementations
\cite{Wilhelm:2018:306,Forster:2020:7381,Wilhelm:2021:1662} in order
to obtain a low-scaling $GW$ algorithm. 
Although the variational instabilities \cite{Merlot:2013:1484,Wirz:2017:4897}
associated to the local fitting approaches are not expected
to be of importance in $GW$ calculations (it is not variational),
we believe care is warranted, especially for the description of core
and semi-core states obtained with the contour-deformation
approach (see below). 

As a consequence, our implementation will use the unconstrained 
global RVF in Coulomb metric. Furthermore, we will assume that 
the auxiliary basis set is orthonormal in the same Coulomb metric, i.e.
\begin{equation}
    \bar{f}_P(r) = \sum\limits_Q \eri{P}{Q}^{-1/2} f_{Q}(r)
\end{equation}
As a result, the four-center ERI
\begin{equation}
    \eri{ai}{bj} = \sum\limits_P \eri{ai}{P}\eri{P}{bj} + \mathcal{O}(\tau_{ai}\tau_{bj})
\end{equation}

\subsection{Spectral Decomposition}
The implementation of the spectral decomposition approach follows the usual transformation\cite{Bauernschmitt:1996:454} of the Casida equations
from a $2N_{occ}N_{vir}$ non-Hermitian eigenvalue problem 
to an $N_{occ}N_{vir}$ Hermitian one of the form
\begin{equation}
    \left(\mathbf{A}-\mathbf{B}\right)^{1/2} \left(\mathbf{A}+\mathbf{B}\right) \left(\mathbf{A}-\mathbf{B}\right)^{1/2} \mathbf{T} = \Omega^2 \mathbf{T}
\end{equation}
where
\begin{equation}
    \mathbf{T} = \left(\mathbf{A}-\mathbf{B}\right)^{-1/2} \left(\mathbf{X}+\mathbf{Y}\right)
\end{equation}
In the RPA, $\left(\mathbf{A}-\mathbf{B}\right)$
is diagonal and consists only of the KS eigenvalue differences
between the occupied and virtual spaces. The elements of 
$(\mathbf{A}+\mathbf{B})$ are obtained using the RVF
approximation as
\begin{equation}
    A_{ia,jb}+B_{ia,jb} = \delta_{ia}\delta_{jb}\left(\epsilon_a -\epsilon_i\right) + 4\sum\limits_P \eri{ia}{P}\eri{P}{jb}
\end{equation}
Once $\mathbf{X}+\mathbf{Y}$ is in hand, the matrix elements
of the screened Coulomb operator can be obtained as
\begin{equation}
    W_{mn,pq}(\omega) = \sum\limits_P  \eri{mn}{P}\eri{P}{pq} + \sum\limits_s w_{mn}^s w_{pq}^s\left(\frac{1}{\omega - \Omega_s + i\eta} - \frac{1}{\omega + \Omega_s - i\eta} \right) 
\end{equation}
where
\begin{equation}
    w_{mn}^s = \sum\limits_{ia} \sum\limits_{P} \eri{mn}{P}\eri{P}{ia} \left( X_{ia}^s + Y_{ia}^s\right)
\end{equation}
Since the largest portion of memory goes into the diagonalization, it is 
almost guaranteed that the two sets of three-center ERIs,
untransformed and contracted with the excitation vectors, can fit in memory.
The final step is to obtain the matrix elements of the self-energy
operator as
\begin{equation}
    \Sigma_{nn}(\omega) = \sum\limits_i\sum\limits_s \frac{w_{in}^s w_{in}^s}{\omega - \epsilon_i + \Omega_s - i\eta} + \sum\limits_a\sum\limits_s \frac{w_{an}^s w_{an}^s}{\omega - \epsilon_a - \Omega_s + i\eta}
\end{equation}

\subsection{Contour Deformation}
The final expressions
are obtained by using the RVF approximation to obtain the matrix elements of the
screened Coulomb operator 
\begin{equation}
    W_{mn\sigma,mn\sigma}(\omega) = \sum\limits_{P,Q}\eri{mn\sigma}{P} \left[ \mathbf{1} - \bm{\Pi}(\omega) \right]_{PQ}^{-1} \eri{Q}{mn\sigma}
    \label{eq:wauxbas}
\end{equation}
and inserting them into Eqs.~\ref{eq:R} and~\ref{eq:I}.
Here, $\bf{\Pi}(\omega)$ corresponds to the representation of 
the polarizability in the auxiliary basis
\begin{equation}
    \Pi_{PQ}(\omega) = \sum\limits_{\sigma}\sum\limits_{i,a} \eri{P}{ia\sigma} \left[ \frac{1}{\omega-\epsilon_{a\sigma} + \epsilon_{i\sigma} + i\eta} + \frac{1}{-\omega-\epsilon_{a\sigma}+\epsilon_{i\sigma} + i\eta} \right] \eri{ia\sigma}{Q}
\end{equation}
and $\bm{\epsilon}(\omega) = \mathbf{1} - \bf{\Pi}(\omega)$ is the
representation of the dielectric matrix in the same basis. The previous
equation follows from the assumption that the irreducible polarizability 
$P$ is equal to the independent particle response function $\chi^{KS}$ or,
equivalently, from the random-phase approximation.

The integral over the imaginary axis, 
\begin{equation}
    I_{nn\sigma}(\omega) = -\frac{1}{2\pi}  \int\limits_{-\infty}^{\infty} d\xi %
    \sum\limits_m \sum\limits_{P,Q} \frac{  \eri{nm\sigma}{P} \left[ \bm{\epsilon}(i\xi)\right]^{-1}_{PQ} \eri{Q}{mn\sigma} }{\omega + i\xi  - \epsilon_{m\sigma} + i\eta \times \mathit{sign}(\epsilon_m - \mu)} 
    \label{eq:inn}
\end{equation}
is computed numerically using a modified Gauss-Legendre grid
\cite{Ren:2012:053020} with  200 points \cite{Golze:2018:4856}. 
The dielectric matrices appearing in Eq.~\ref{eq:inn}, 
depend on purely imaginary frequencies, are Hermitian 
positive-definite \cite{Holzer:2019:204116} and do not depend 
on the particular states $\phi_m$ or $\phi_n$.
As a consequence, all screened Coulomb matrix elements
\begin{equation}
    W_{mn\sigma,mn\sigma}(i\xi) = \sum\limits_{P,Q} \eri{nm\sigma}{P} \left[ \bm{\epsilon}(i\xi)\right]^{-1}_{PQ} \eri{Q}{mn\sigma}
\end{equation}
can be pre-computed in a very 
efficient manner. In the actual implementation we use the
rectangular full packed \cite{Gustavson:2010:1} subroutines 
implemented in LAPACK \cite{LAPACK}.

The residue term, 
\begin{eqnarray}
    R_{nn\sigma}(\omega) = &-& \sum\limits_i \sum\limits_{P,Q} \eri{ni\sigma}{P} \left[ \bm{\epsilon}(\varepsilon_i - \omega + i\eta)\right]^{-1}_{PQ} \eri{Q}{in\sigma} \theta(\varepsilon_i -\omega) \nonumber \\
    &+& \sum\limits_a \sum\limits_{P,Q} \eri{na\sigma}{P} \left[ \bm{\epsilon}(\varepsilon_a - \omega - i\eta)\right]^{-1}_{PQ} \eri{Q}{an\sigma}\theta(\omega - \varepsilon_a)
    \label{eq:rnn}
\end{eqnarray}
 requires a slightly different approach since the dynamic dielectric 
 matrices depend both on the real frequencies $\omega$ as well as on 
 the eigenstates energies. The $\mathcal{O}(N^5)$ scaling of this 
step--$N_{\mathrm{occ}}^2 \times N_{\mathrm{vir}} \times N_{\mathrm{aux}}^2$
for the core states--is given by the explicit computation 
of such matrices. Avoiding the explicit computation of
the dielectric matrices thus becomes very important.

In order to do so, we use the minimal residual (MINRES)
\cite{Paige:1975:617} or the Eirola-Nevanlinna (EN)
\cite{Eirola:1989:511} iterative solvers for indefinite matrices.
The EN solver was previously used by one of us to solve
linear equation systems \cite{MejiaRodriguez:2015:1493} closely 
related to Eq.~\ref{eq:rnn}. However, the
equations in CD-$GW$ are symmetric, which allows for
the use of the more efficient MINRES solver. The use of either
iterative solver reduces the scaling by one order of magnitude,
provided that the number of steps needed to reach the 
solution is much smaller than the number of auxiliary functions.

The actual implementation defaults to MINRES with
a maximum of 25 iterations and a $5\times 10^{-5}$ 
convergence threshold on the residual norm.
If the norm is still larger than
$10^{-4}$ after 25 iterations, an indication of
very slow convergence, the dielectric
matrix is actually built and the solution is obtained
via the Bunch-Kaufman factorization \cite{Bunch:1977:163}.
This switch can be adjusted depending on the size of the
ERI tensor in order to ensure that the fastest approach
is always used.

The pseudocode for the CD-$GW$ implementation is shown in
Algorithm \ref{algorithm}. Note that, as stated above, the computation
of the matrices $\left[\mathbf{1} - \bm{\Pi}(\omega)\right]^{-1}$ is
avoided as much as possible by using MINRES to
obtain $\left[\mathbf{1} - \bm{\Pi}(\omega)\right]^{-1} \mathbf{R}_{mn}$
without even explicitly building the $\bm{\Pi}$ matrices.
A very important aspect in the overall efficiency of the code
is the $\omega$ step described in the next subsection.

\begin{algorithm}
\SetAlgoLined
\DontPrintSemicolon
Compute DFT ground state to calculate eigenvectors $\mathbf{C}$ and eigenvalues $\bm{\varepsilon}$\;
\BlankLine
\myInit{
Read $\mathbf{C}$ and $\bm{\varepsilon}$\;
Compute  and transform $\eri{\mu\nu}{P} \rightarrow \eri{m\nu}{P} \rightarrow \eri{mn}{P} = E_{mn,P}$ for all $P$'s assigned to the processor\;
Redistribute ERIs by MO pairs\;
Orthonormalize ERIs as $\mathbf{R} = \mathbf{L}^{-1} \mathbf{E} $\;
Compute and transform $\left( \mu | v_{xc} \, \nu \right) \rightarrow \left( n | v_{xc} n \right) = V_{nn}^{xc}$\;
Compute $\Sigma_{n,n}^x = \left( n | \Sigma^x | n \right) = -\sum_i^{occ}\sum_k | \left( k || i n \right)|^2$\;
Set $\Sigma_{n,n}^{old} = V_{nn}^{xc}$\;
}\;
\myCDGW{
  \BlankLine 
  \For{$iter=0$,$maxeviter$}{
    Compute energy differences $\omega_{ia} = \epsilon_i - \epsilon_a$\;
    \ForAll{$\omega' \in$ imaginary grid}{
       Compute $W_{mn,mn}(\mathrm{i}\omega') = \mathbf{R}_{mn}^T\; \left[  \mathbf{1} - \bm{\Pi}(\mathrm{i} \omega') \right]^{-1} \; \mathbf{R}_{mn}$\;
    }\;
    \For{$n\in$ quasiparticle energies requested}{
      \While{not converged}{
        Update $\Sigma_{n,n} \gets -\sum_{g} \sum_m^{all} z_g W_{mn,mn}(\mathrm{i}\omega'_g) \; \left(\omega - \epsilon_m\right) / \left( \omega_g'^2 + \left(\omega - \epsilon_m\right)^2 \right)$ and its derivative\;
        \ForAll{$|\epsilon_m| \leqslant |\omega|$}{
          Set $f_m = \mathrm{sgn}(\omega) \; \theta\left( \mathrm{sgn}(\omega) \, (\omega - \epsilon_m) \right)$\;
          Update $\Sigma_{n,n}^c(\omega) \gets f_m \mathbf{R}_{mn}^T \left[ \mathbf{1} - \bm{\Pi}(\epsilon_m - \omega) \right]^{-1} \mathbf{R}_{mn}$ and its derivative\;
        }\;
        Compute $\Sigma_{n,n}^c(\omega)$, $ \partial_\omega \Sigma_{n,n}^c(\omega)$ \;
        Update $\omega$ according to solver\;
      }\;
      Set $\epsilon_n^{QP} = \omega$\;
      Set $\Sigma_{n,n}^{old} = \Sigma(\omega)$\;
    }\;
    Update $\bm{\epsilon} = \bm{\epsilon}^{QP}$\; 
  }\;
}\;
\caption{Contour Deformation (CD) $GW$ \label{algorithm}}
\end{algorithm}

\subsection{Solution of the Quasiparticle Equations}

In the $GW$ approximation, the exchange-correlation operator of the underlying mean-field theory gets replaced by the non-locala and dynamical self-energy operator. Therefore, the corrections to
the mean-field orbital energies $\varepsilon_{k\sigma}$ are given by:
\begin{equation}
    \epsilon_{k\sigma}^{QP} = \epsilon_{k\sigma} + \operatorname{Re}\Sigma_{kk\sigma} ( \epsilon_{k\sigma}^{QP}) - V_{kk\sigma}^{xc} \label{eqn:qpeqn}
\end{equation}
These so-called quasiparticle equations must be solved self-consistently for a given $G$ and $W$. 

Eq.~\ref{eqn:qpeqn} can be solved using one of several
approaches. A \emph{graphical} solution can be obtained 
by computing the self-energy on a fine grid of real
frequencies in the region were the solution is expected.
An \emph{iterative} optimizer, like the the Newton or
Nelder-Mead algorithms \cite{NelderMead} among others, 
might also be used to find one of the--possibly many--fixed-points 
of Eq.~\ref{eqn:qpeqn}. Finally, a so-called 
\emph{linearized approximation}, which
is formally equivalent to one step of Newton's method,
has also been successfully used for valence states,
but has important failures for core or
semi-core states \cite{Golze:2018:4856}.

For the spectral decomposition approach, the most
computationally demanding parts are the
diagonalization of the Casida-like Eq.~\ref{eqn:rpa}
and the subsequent contraction of the ERIs
with the obtained eigenvectors, with 
$\mathcal{O}(N^6)$ and $\mathcal{O}(N^5)$ formal
scalings, respectively. Fortunately, both steps
are frequency-independent.

In contrast, the most demanding tasks of the
contour-deformation approach are frequency-dependent.
In particular, building and inverting the dielectric matrices
$\mathbf{1} - \bm{\Pi}(\xi)$ for all residues 
scale as $\mathcal{O}(N^5)$ for core states.

Different solvers are used for the
spectral decomposition and the contour-deformation approaches
due to the aforementioned intrinsic computational differences.
A graphical solver is used by default when the user
requests the spectral decomposition approach, whereas
an iterative solver based on Newton's method is used
when the contour-deformation approach is requested.

In order to further minimize the number of frequency dependent self-energies ($\Sigma(\omega)$)
evaluated in CD-$GW$, the Newton method has been complemented
as follows. The first step is always an scaled residual
\begin{equation}
    f(\xi) = \varepsilon_{k\sigma} + \operatorname{Re}\Sigma_{kk\sigma} (\xi) - V_{kk\sigma}^{xc} - \xi
\end{equation}
in order to avoid computing the derivative
$\partial_\xi \operatorname{Re}\Sigma_{kk\sigma} (\xi)$ near
a pole of $G_{KS}$. Further steps are decided depending on
the value the value of the so-called renormalization
factor
\begin{equation}
    Z = -\left(\frac{\partial f(\xi)}{\partial \xi}\right)^{-1}
\end{equation}
and whether the solution has been bracketed or not. A bracket is 
found whenever $f(\xi)$ changes sign, or when $f(\xi)$ keeps the
same sign but its magnitude increases between consecutive steps (this is often a sign of a skipped solution).

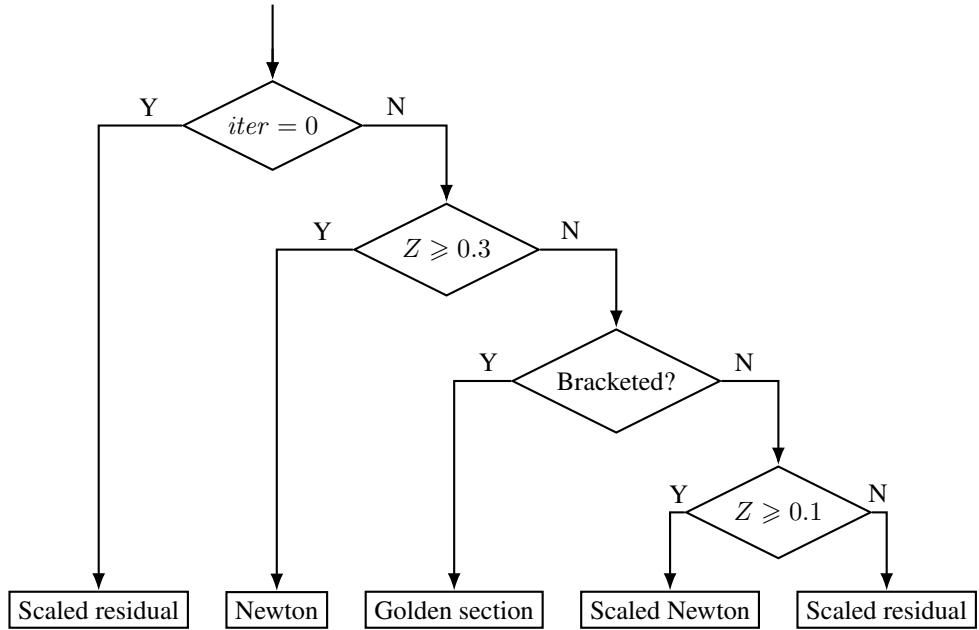
\begin{figure}
\centering
\forestset{
  default preamble={
    before typesetting nodes={
      !r.replace by={[, coordinate, append]}
    },
    where n children=0{
      tier=word,
    }{
      diamond, aspect=2,
    },
    where level=0{}{
      if n=1{
        edge label={node[pos=.2, above] {Y}},
      }{
        edge label={node[pos=.2, above] {N}},
      }
    },
    for tree={
      edge+={thick, -Latex},
      math content,
      s sep'+=0.25cm,
      draw,
      thick,
      edge path'={ (!u) -| (.parent)},
    }
  }
}
{\footnotesize
\begin{forest}
[{\mathit{iter}=0}, tikz={\draw[{Latex}-, thick] (.north) --++ (0,1);}
    [ \text{Scaled residual} ]
    [ Z \geqslant 0.3
        [ \text{Newton} ]
        [ \text{Bracketed?}
            [ \text{Golden section} ]
            [  Z \geqslant 0.1 
                [ \text{Scaled Newton} ]
                [ \text{Scaled residual} ]
            ]
        ]
    ]
]
\end{forest}
}
\caption{Iterative solver step selection tree.}
\label{fig:steptree}
\end{figure}

Figure \ref{fig:steptree} shows the decision tree used to
obtain the step size and direction. The scaling factor
for the scaled Newton step is set to $0.7$, while the scaling
factor for the scaled residual step adapts to the actual 
magnitude of the residual but is never larger than $0.1$.
We have noticed that in many cases where $Z<0.3$, the Newton algorithm leads to slow convergence of the quasiparticle equations. One way to alleviate this issue is to switch to the golden-section search when the solution has been bracketed and $Z<0.3$, as shown in Figure \ref{fig:steptree}. 

In order to further minimize the number of frequencies probed,
we have added an additional layer. In a large molecule with many nuclei 
of the same type, the quasiparticle energies will tend to be clustered
around several small regions. Once a solution for one state inside
each cluster is found, a very good initial guess for the rest of the
states of that cluster is in hand. This initial guess can be used
either to define a relevant region to search for a graphical solution, 
or to start the iterative solver for a new quasiparticle within the
cluster.

\section{Results}
\label{res}
\subsection{Valence Spectra}
In order to assess the correctness of the current
implementation, $G_0W_0$@PBE/def2-QZVP HOMO and LUMO energies 
from the GW100 benchmark set \cite{vanSetten:2015:5665}
were compared to reference values obtained from the
GW100 GitHub repository \cite{GW100GitHub}.

\begin{figure}
    \centering
    \includegraphics[width=0.75\textwidth]{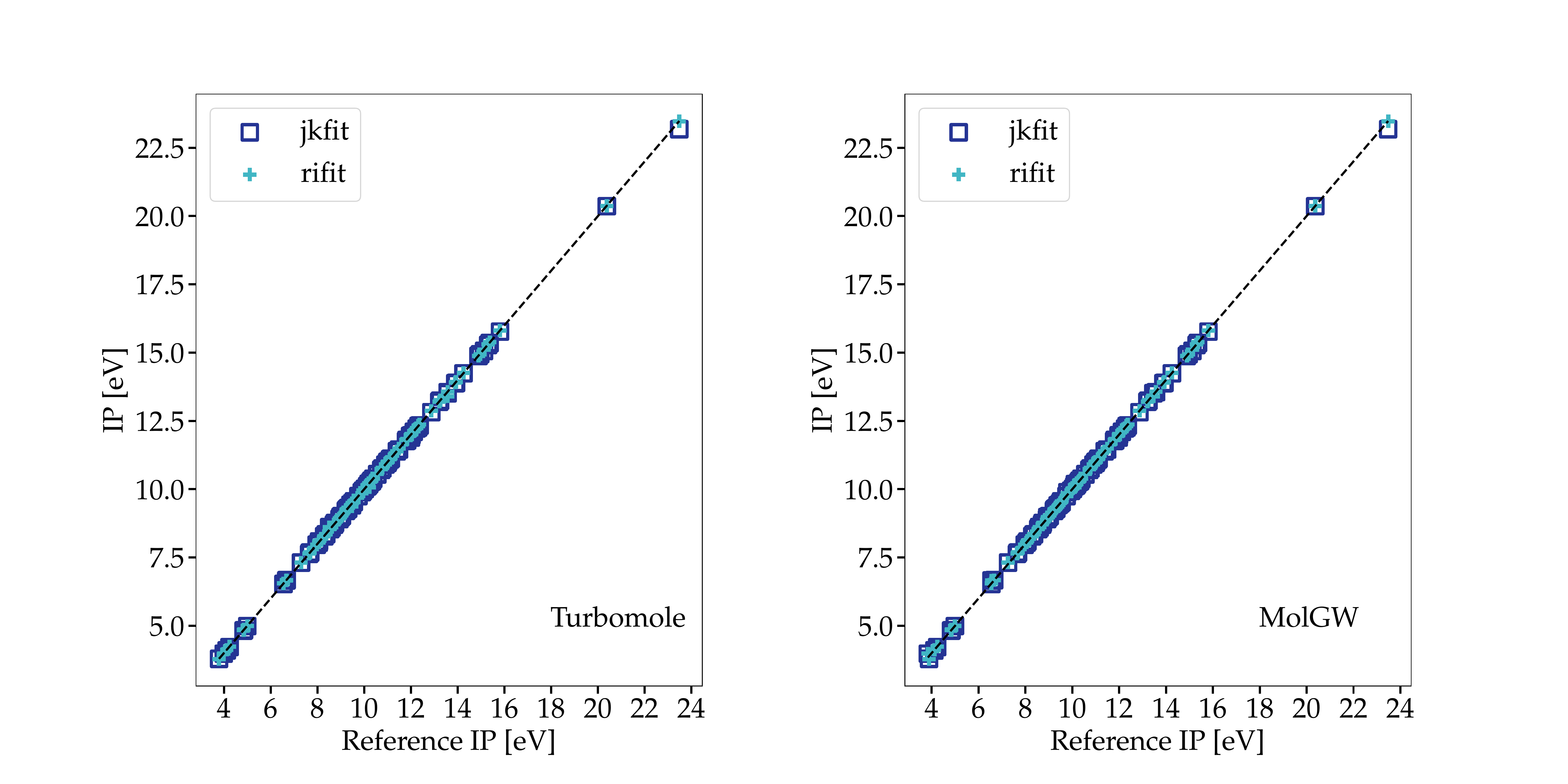}
    \caption{Comparison between the $G_0W_0$@PBE vertical ionization potentials computed with the current implementation and two other codes. Results from Turbomole and MolGW were obtained from the GW100 GitHub repository \cite{GW100GitHub} and correspond to the entries ``TM v7.0 def2-QZVP RIK'' and ``Mv2.A def2-QZVP cc-pVQZ-RI (first peak)''. All values in eV.}
    \label{fig:gw100_homo}
\end{figure}

Figure \ref{fig:gw100_homo} shows correlation plots for all
100 vertical ionization potentials using two different auxiliary
basis sets: def2-universal-jkfit \cite{JKFIT} (JKFIT) and 
def2-qzvp-rifit\cite{RIFIT} (RIFIT). The former
is smaller and designed for Coulomb and exact-exchange fitting,
while the latter is designed for correlation calculations.
The same auxiliary basis set was used for the ground-state calculation
as well as the subsequent $GW$ calculation.
The results obtained in this work with the RIFIT auxiliary set are on top of 
the Turbomole reference using the same auxiliary set published in the GW100 GitHub repository \cite{GW100GitHub}. We also show MolGW results obtained with
an automatically generated auxiliary set and published in the same
repository.

The JKFIT auxiliary set yields almost the same accuracy as the larger 
RIFIT one except for the helium atom, where a large 0.30 eV deviation 
can be observed. This deviation was already noted in the original 
GW100 manuscript (Ref. \citenum{vanSetten:2015:5665}), with an undisclosed
auxiliary set. This issue is already present in the ground-state 
KS reference, where the JKFIT occupied eigenvalue deviates by the
same amount from a calculation without RVF.

Overall, if the analytical full-frequency results from Turbomole are 
taken as reference, the JKFIT auxiliary basis leads to a
mean absolute error (MAE) of 23 meV, while the RIFIT cuts the error
in half to only 10 meV (compare to 13 meV MAE for MolGW).

\begin{figure}
    \centering
    \includegraphics[width=0.75\textwidth]{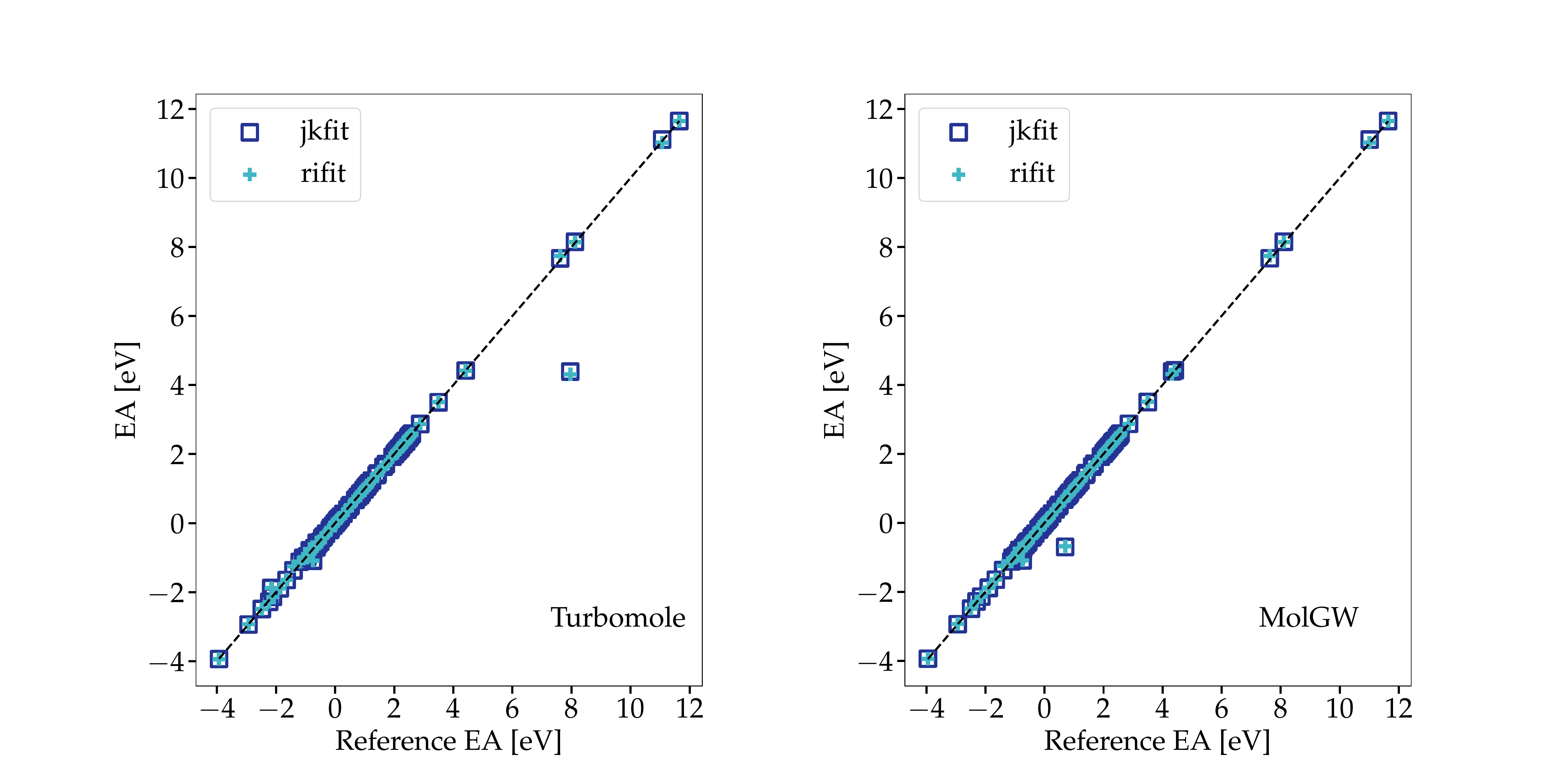}
    \caption{Comparison between the $G_0W_0$@PBE vertical electron affinities computed with the current implementation and two other codes. Results from Turbomole were taken from Ref. \cite{vanSetten:2015:5665}. MolGW results were obtained from the GW100 GitHub repository \cite{GW100GitHub} and correspond to the entry ``Mv2.B\_def2-QZVP\_auto\_firstpeak''. All values in eV.}
    \label{fig:gw100_lumo}
\end{figure}

Similar to the ionization potential plots, Figure \ref{fig:gw100_lumo}
shows correlation plots for the computed vertical electron affinities.
The outliers seen in Figure \ref{fig:gw100_lumo} correspond to the
xenon atom in the Turbomole plot, and to the fluorine dimer in the
MolGW plot. Interestingly, Turbomole and MolGW also differ
between each other in these two cases. The MAEs worsen to 58 meV,
66 meV, and 73 meV, for JKFIT, RIFIT and MolGW, respectively.

\subsection{Core Spectra}
\begin{figure}
    \centering
    \includegraphics[width=0.6\textwidth]{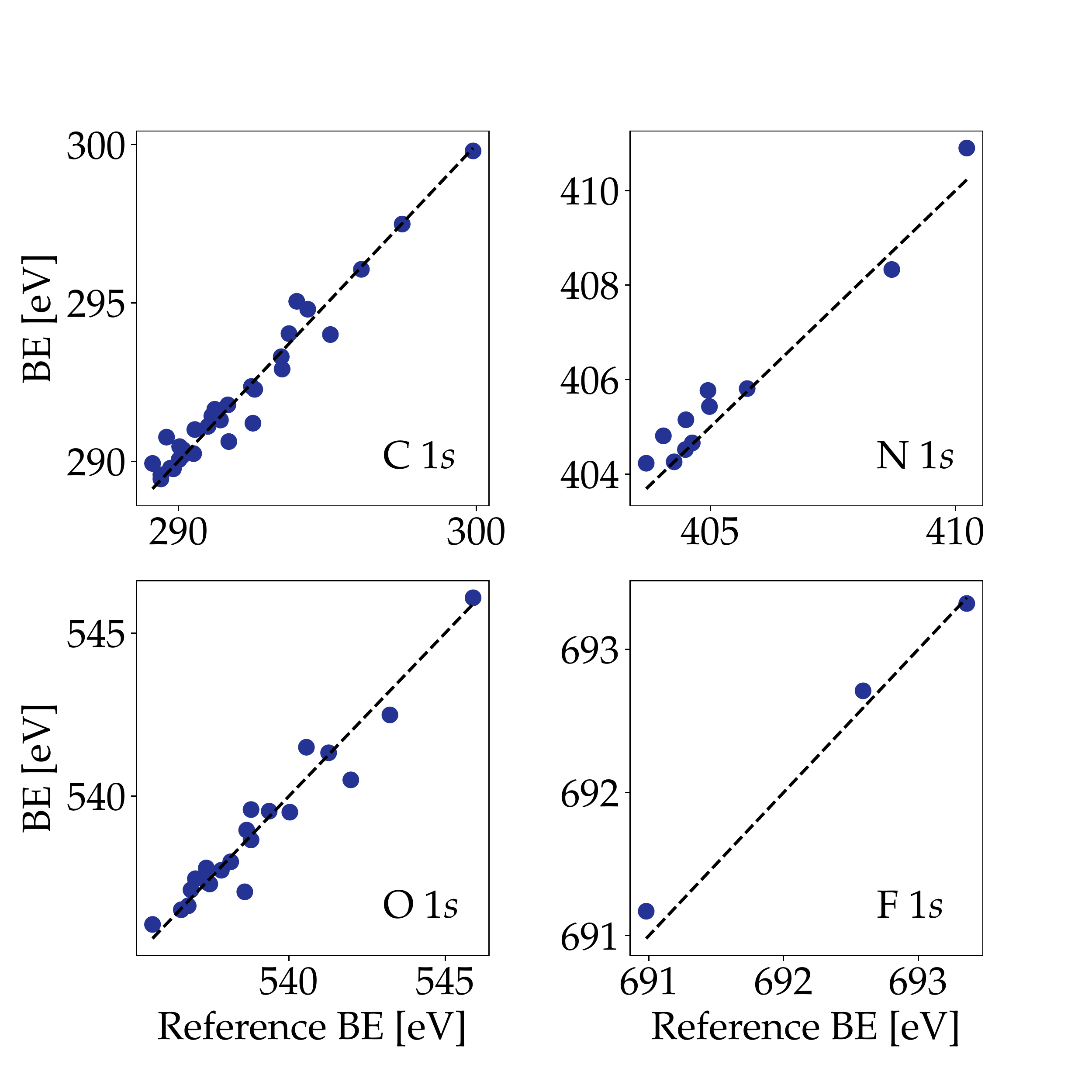}
    \caption{Correlation plot of the CORE65 ev$GW_0$@PBE/cc-pvtz binding energies obtained with the current implementation and from Ref. \citenum{Golze:2020:1840}. The dashed lines indicate an exact match.}
    \label{fig:core65_pbe}
\end{figure}

\begin{figure}
    \centering
    \includegraphics[width=0.6\textwidth]{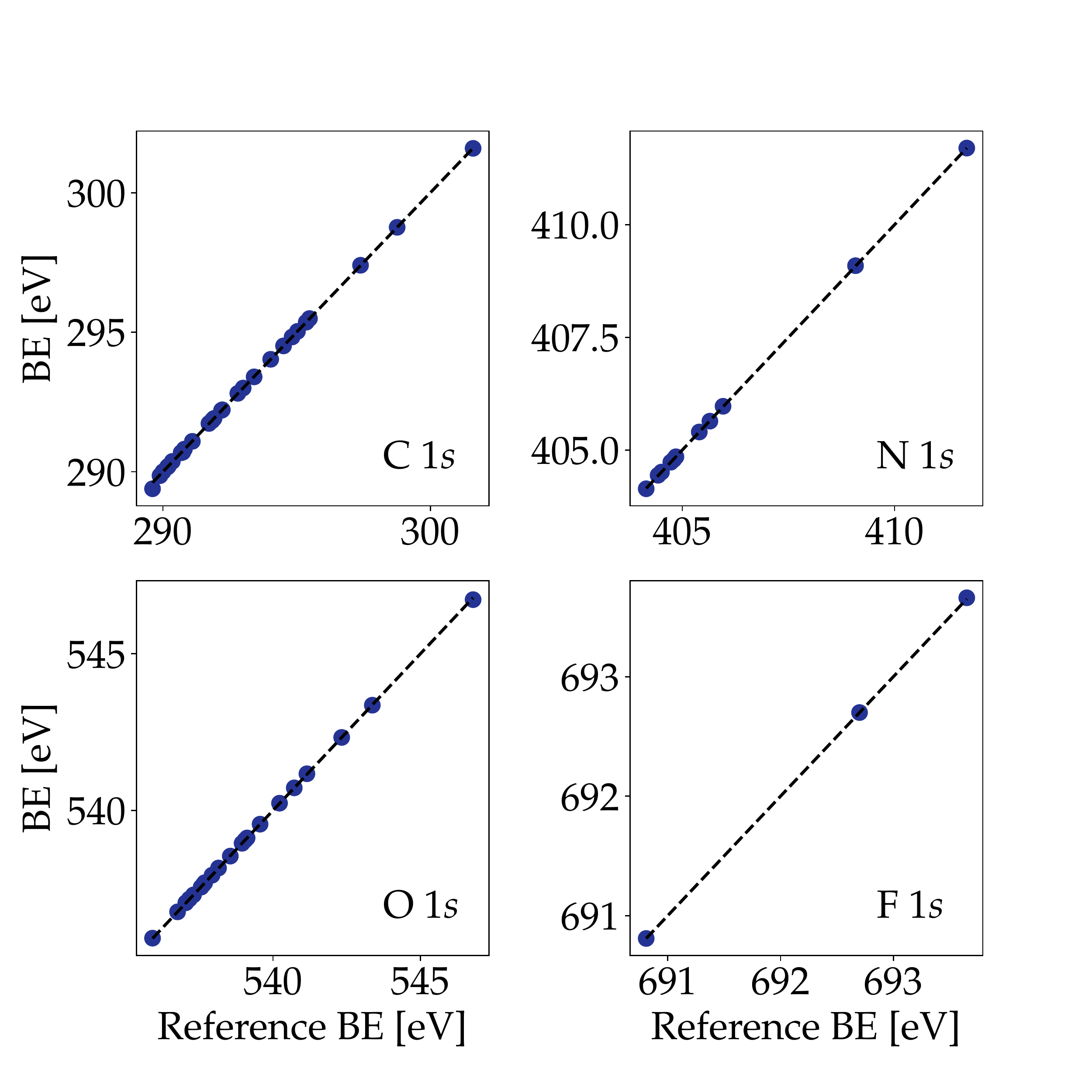}
    \caption{Correlation plot of the CORE65 $G_0W_0$@PBEh/cc-pvtz binding energies obtained with the current implementation and from Ref. \citenum{Golze:2020:1840}. The dashed lines indicate an exact match.}
    \label{fig:core65_pbeh}
\end{figure}

The accuracy of the core-level was assessed using the
CORE65 benchmark set from Golze \emph{et al.}
\cite{Golze:2020:1840}. The CORE65 set contains 
32 small inorganic and organic molecules with up to 
14 atoms. We present comparisons to the
non-relativistic results at the ev$GW_0$@PBE,
$G_0W_0$@PBEh
(see Supporting Information of Ref. \citenum{Golze:2020:1840}),
$G_0W_0$@PBE0, and ev$GW$@PBE0 levels of theory. All
calculations used the cc-pvtz basis set and the
JKFIT auxiliary basis. Here,
PBEh \cite{Atalla:2013:165122} refers to the 
hybrid functional with
45\% Hartree-Fock (HF) exchange, and PBE0 \cite{Perdew:1996:9982,Adamo:1999:6158} 
to the functional with 25\% HF exchange.
Figures \ref{fig:core65_pbe} and \ref{fig:core65_pbeh} show
correlation plots for the aforementioned cases (see
Supporting Information for individual results.)

The correlation shown for the results obtained using the PBE GGA functional is not close to linearity.
This might be explained by the 
difficulty to find the exact same solution
along the whole ev$GW_0$ self-consistent cycle 
between the two different solvers, as
there are many very close solutions for the quasiparticle
equations of such states\cite{Golze:2018:4856,Golze:2020:1840}.
In fact, our experience indicates that small energy differences ($\sim\mu$ eV) might lead 
our own solver to different solutions, and that these differences 
sometimes disappear during the ev$GW_0$ and ev$G_0W_0$ cycles.
Nevertheless, the perfect correlation seen in the
PBEh core results, as well as the
very good agreement in the valence region, reassures us 
about the soundness of our implementation. 
The large fraction of HF exchange in the hybrid PBEh functional facilitates the identification
of a single solution for each quasiparticle equation
\cite{Golze:2018:4856,Golze:2020:1840}, 
leading to the expected match between codes.

Including 25\% HF exchange, as in the PBE0 functional, does not 
completely resolve the multiple solution issues in the core region,
but it does help to reduce the number of tightly clustered peaks.
Another drawback of using a low percentage of HF exchange is that
the one-shot $G_0W_0$ binding energies are still not good enough to
avoid doing the more expensive ev$GW$ procedure (see Supporting Information), 
while the SCF step is as expensive as any other hybrid functional with 
larger fraction of HF exchange.

\subsubsection{ESCA Molecule}
Ethyl trifluoroacetate, also known as the ESCA molecule \cite{Siegbahn:1967},
shows extreme chemical shifts in its carbon 1{\it s} (C 1{\it s}) binding energies and 
has been an important reference system since the dawn of photoelectron
spectroscopy \cite{Travnikova:2012:191}. Previous computational 
studies involving the C 1{\it s} binding energies of the ESCA molecule 
have been performed using the $\Delta$SCF method \cite{Travnikova:2012:191,vandenBossche:2014:034706,Delesma:2017:169,Travnikova:2019:7619,Klein:2021:154005}, 
but we could not find any core-level $GW$ studies reported for this molecule.

\begin{figure}
    \centering
    \includegraphics[width=0.3\textwidth]{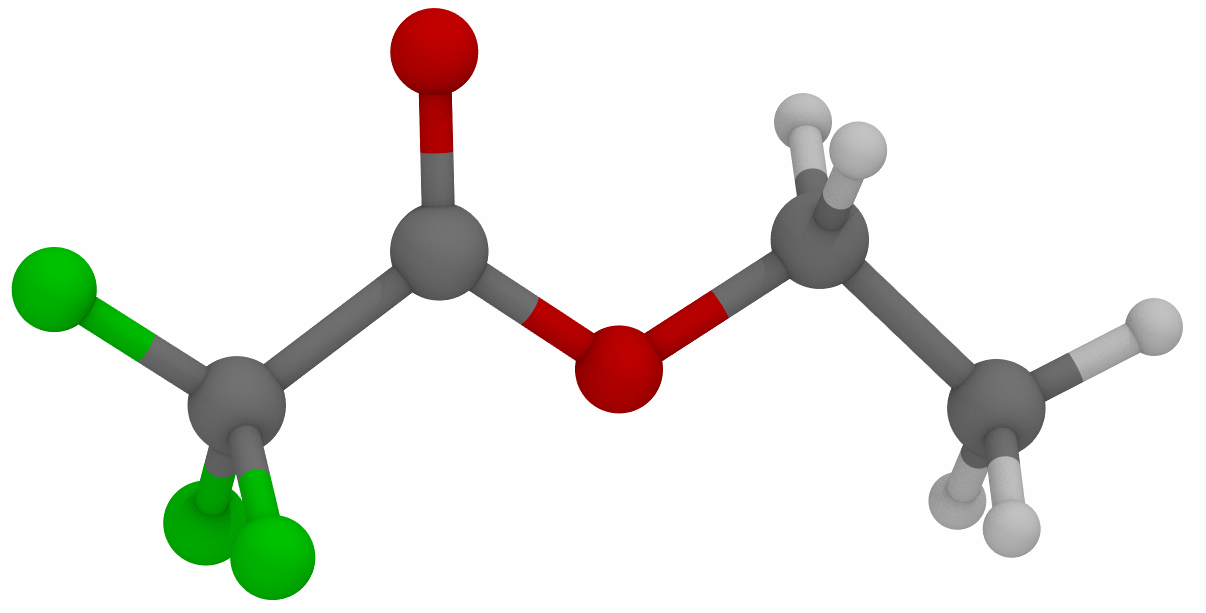}\hspace{1.5cm}
    \includegraphics[width=0.3\textwidth]{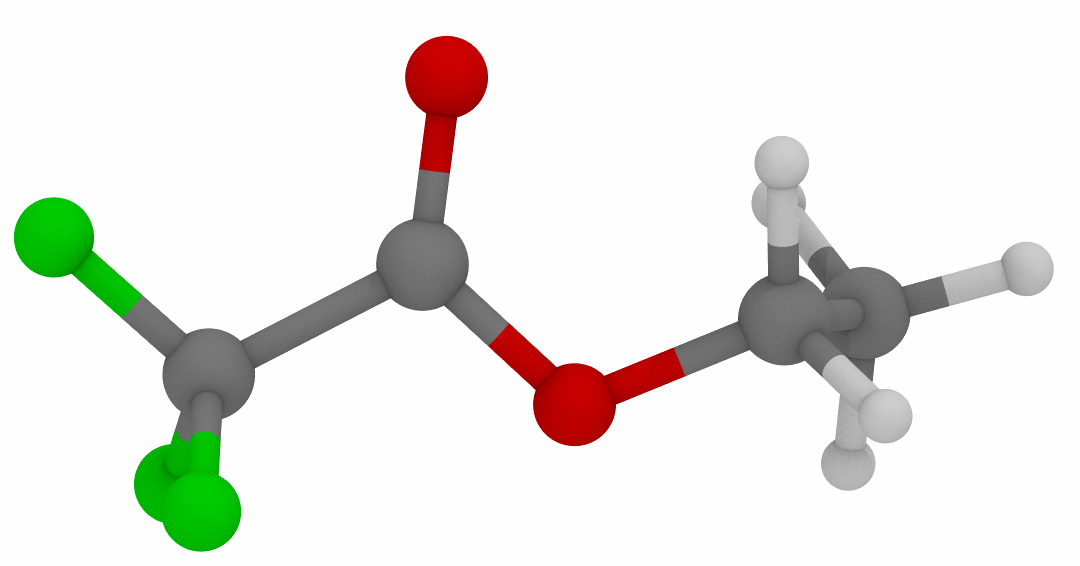}
    \caption{Ethyl trifluoroacetate in the anti-anti conformation (left) and anti-gauche conformation (right).}
    \label{fig:esca}
\end{figure}

\begin{table}
    \centering
    \caption{Core-level binding energies of ethyl trifluoroacetate. Experimental results with respect to the vacuum level \cite{Travnikova:2012:191}. All values in eV.\label{tab:esca}}
    \begin{tabular}{c c c c c c }
    \multirow{2}{*}{C 1{\it s} peak} & \multirow{2}{*}{Experimental} & ev$GW_0$ & ev$GW_0$  & ev$GW_0$ & $G_0W_0$ \\
              &              & PBE    & r$^2$SCAN-L   & r$^2$SCAN & PBEh    \\\hline
    \ce{CH3}  &  291.47      & 289.03 & 290.62 & 291.79    & 291.46    \\
    \ce{CH2}  &  293.19      & 291.70 & 292.87 & 293.48    & 293.42    \\
    \ce{CO2}  &  295.80      & 293.95 & 295.24 & 296.23    & 296.43    \\
    \ce{CF3}  &  298.93      & 297.48 & 298.55 & 299.41    & 299.64    \\\hline 
    \end{tabular}
\end{table}

Table \ref{tab:esca} shows the absolute binding energies of
the four C 1{\it s} states of the ESCA molecule. The experimental
results were taken from the high-resolution photoelectron 
spectrum of Travnikova and cols.\cite{Travnikova:2012:191}
The computed binding energies are weighted averages of the two
conformations found in a gas-phase electron diffraction spectrum\cite{Lestard:2010:1357} and shown in Figure \ref{fig:esca}.

All calculations used the experimental geometries, the
pcSseg-3 quadruple-$\zeta$ basis set from Jensen\cite{Jensen:2008:719}
and the def2-universal-jkfit auxiliary basis set. A larger
auxiliary basis set adapted to the pcSseg-3 basis using an
automatic generator\cite{Stoychev:2017:554} does not change the 
quality of the results.

Previously, Fouda and Besley found that the pcSseg-$n$ family
converges faster to the basis set limit in DFT calculations 
of core-electron spectroscopies.\cite{Fouda:2018:6} We also 
found that the pcSseg-$n$ family describes both core and valence 
ionizations more uniformly than the very recent ccX-$n$Z 
family\cite{Ambroise:2021}. 

Four density functional approximations--
PBE, r$^2$SCAN\cite{Furness:2020:8208}, r$^2$SCAN-L\cite{MejiaRodriguez:2020:121109}, 
and PBEh--were used to evaluate the starting points obtained 
from the three major exchange-correlation approximation families.
No relativistic corrections were included.

Clearly, the ev$GW_0$@r$^2$SCAN method yields the best overall
agreement with experiment, with $G_0W_0$@PBEh results
closely following.
Even so, $G_0W_0$@PBEh is the method of choice because it leads
to single solutions for the C 1{\it s} states.
At the other end, we find the binding energies obtained 
with ev$GW_0$@r$^2$SCAN-L and ev$GW_0$@PBE
underbind the C 1{\it s} states but, as expected, ev$GW_0$@r$^2$SCAN-L
energies fall in between those of ev$GW_0$PBE and ev$GW_0$@r$^2$SCAN, respectively. Comparisons with the $\Delta$-SCF method are presented in the Supporting Information.


\section{Parallel Performance}
\label{par-perf}
\begin{figure}
    \centering
    \begin{subfigure}[b]{0.3\textwidth}
        \centering
        \includegraphics[width=0.7\textwidth]{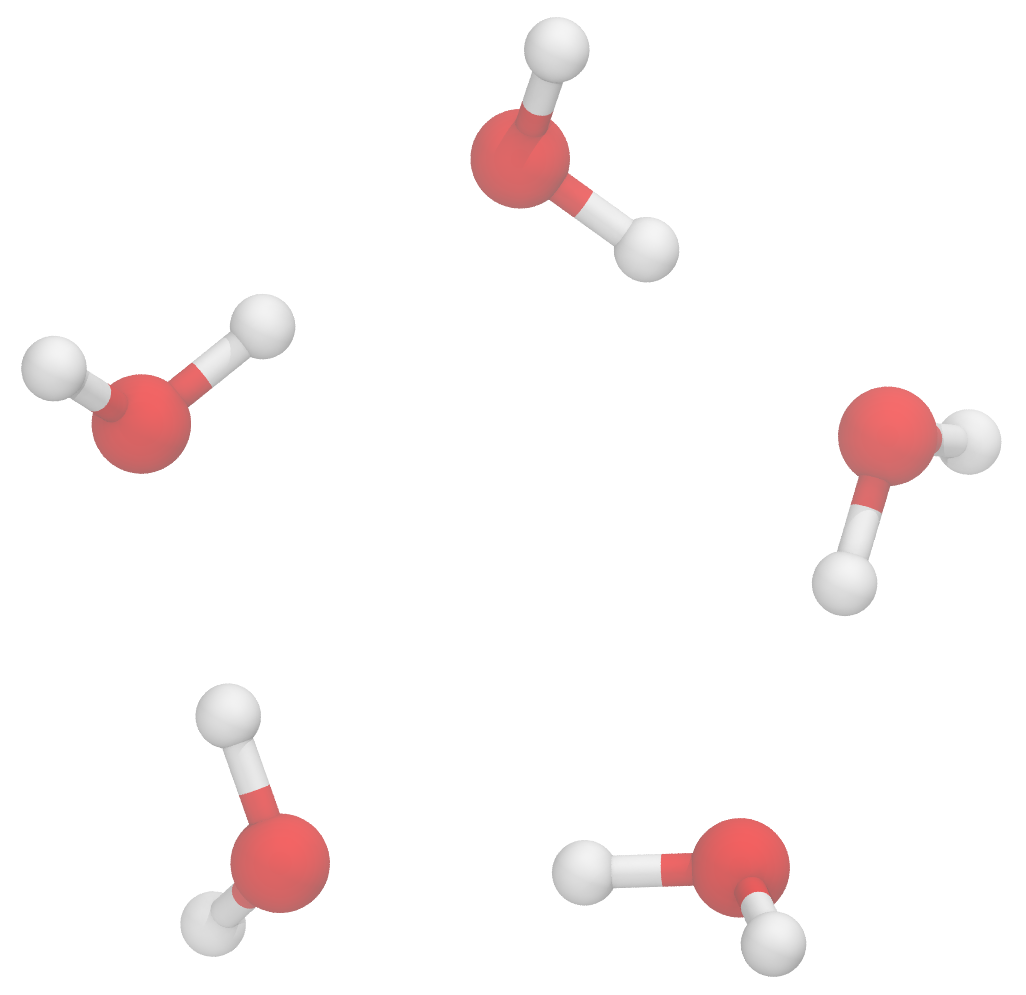}
        \caption{\ce{(H2O)_5}}
    \end{subfigure}
    \hspace{0.5cm}
    \begin{subfigure}[b]{0.3\textwidth}
        \centering
        \includegraphics[width=0.7\textwidth]{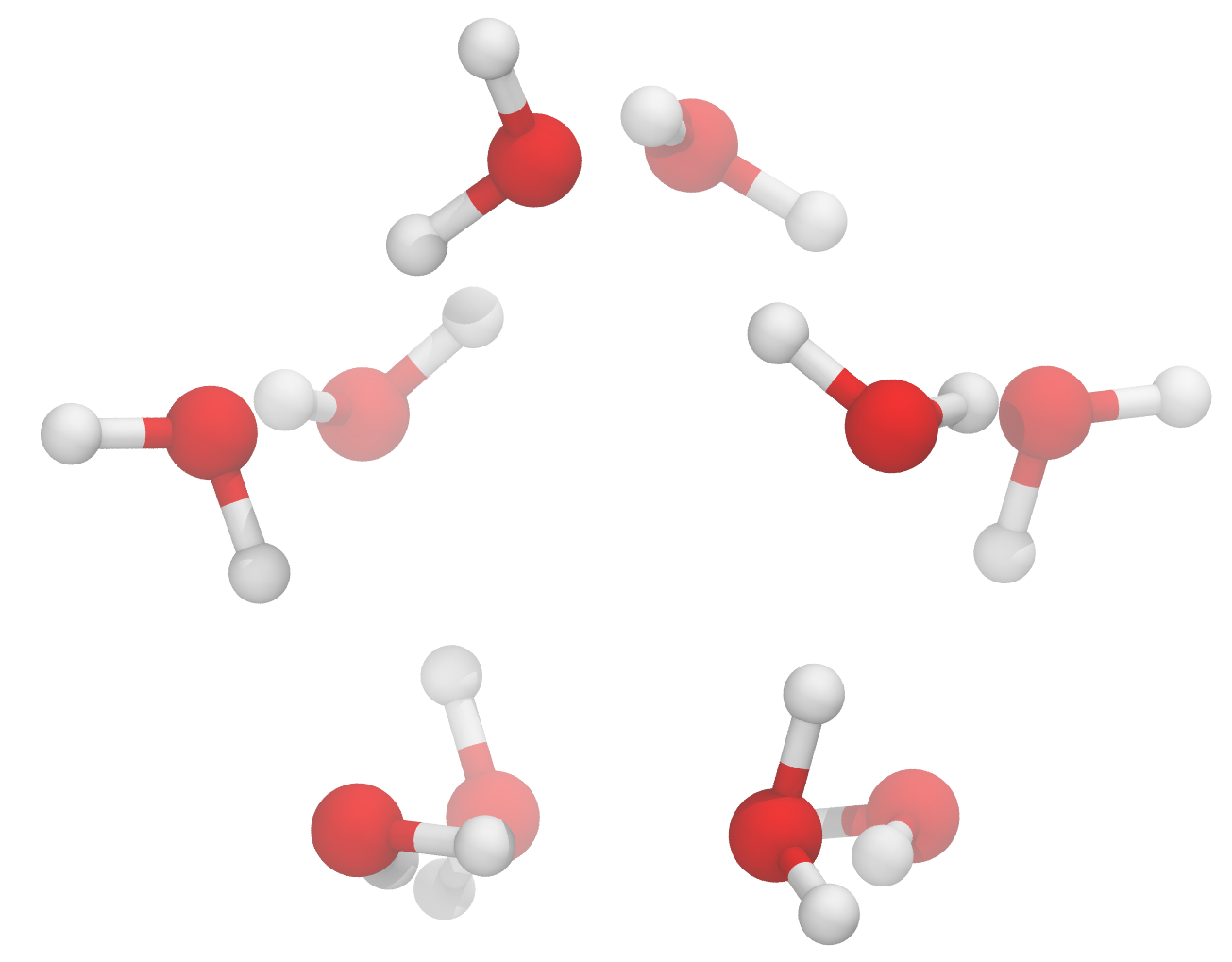}
        \caption{\ce{(H2O)10}}
    \end{subfigure}\\
    \begin{subfigure}[b]{0.3\textwidth}
        \centering
        \includegraphics[width=0.7\textwidth]{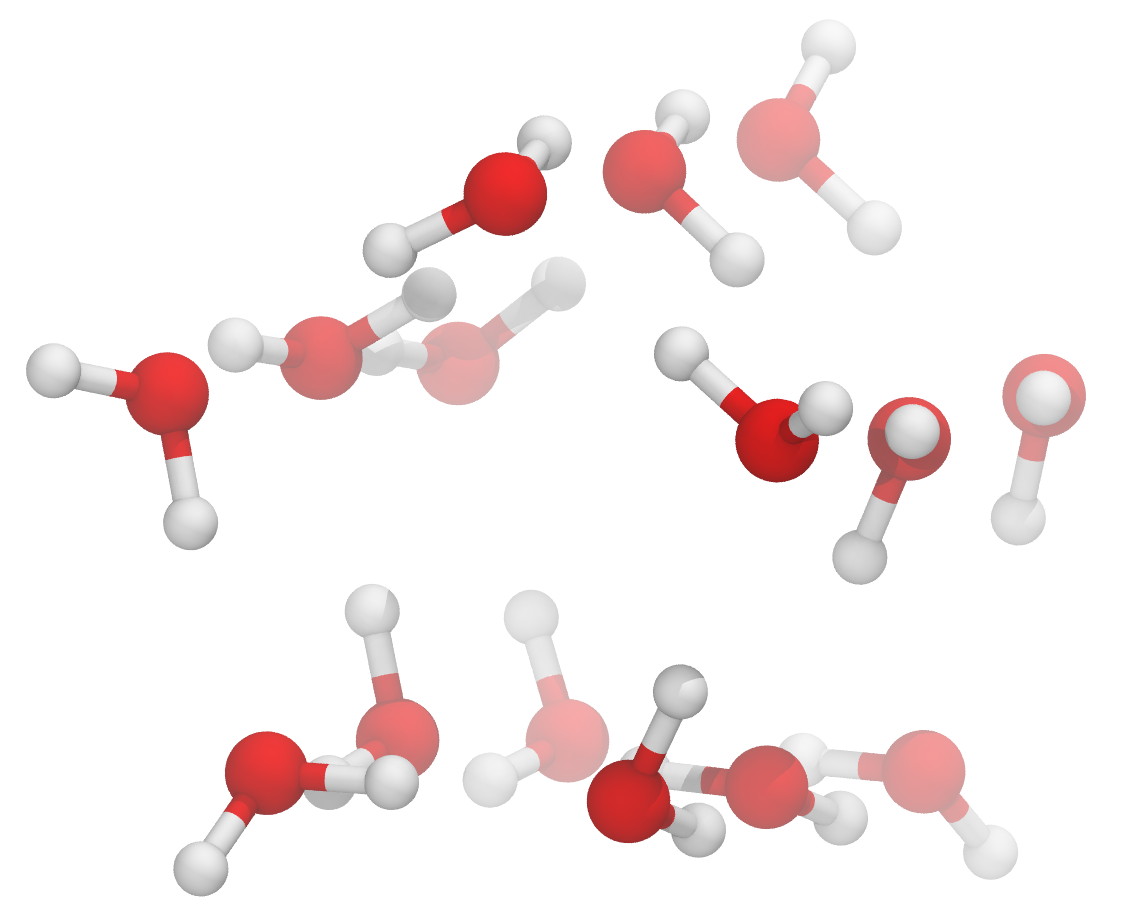}
        \caption{\ce{(H2O)15}}
    \end{subfigure}
    \hspace{0.5cm}
    \begin{subfigure}[b]{0.3\textwidth}
        \centering
        \includegraphics[width=0.7\textwidth]{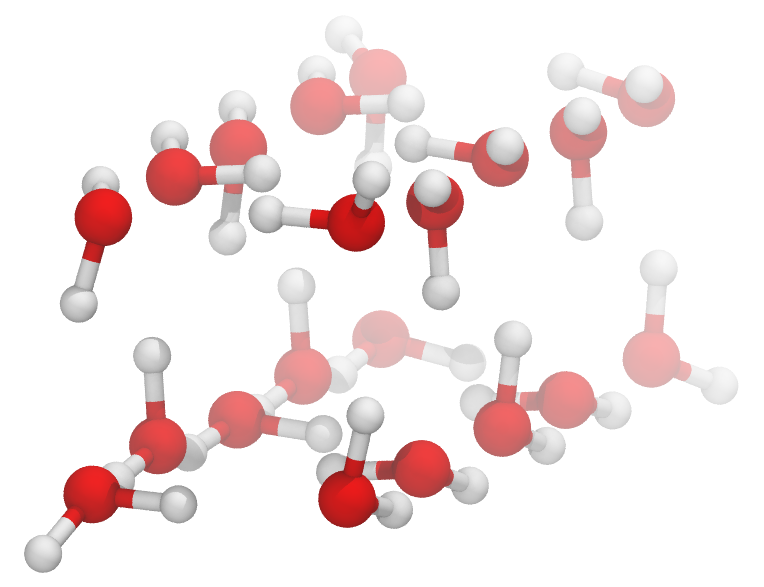}
        \caption{\ce{(H2O)_20}}
    \end{subfigure}    
    \caption{Water cluster structures used to evaluate the computational performance of the CD-$G_0W_0$ method. }
    \label{fig:water}
\end{figure}

The parallel performance and computational scaling of 
the CD-$G_0W_0$ implementation are assessed using water 
clusters with 5, 10, 15, and 20 molecules published in the
Cambridge Cluster Database\cite{Maheshwary:2001:10525}. The molecular
structures of these clusters are shown in Figure \ref{fig:water}.
All calculations presented in this subsection use the
def2-QZVP/def2-universal-jkfit combination of orbital
and auxiliary basis sets without exploiting 
molecular symmetry.

Each computational node used consists of 2 18-core sockets
for a total of 36 Intel Xeon Gold 6254 cores per node with 
up to 384 GB of memory. However, we used a maximum of 32
cores per node. 

\begin{figure}
    \centering
    \includegraphics[width=0.75\textwidth]{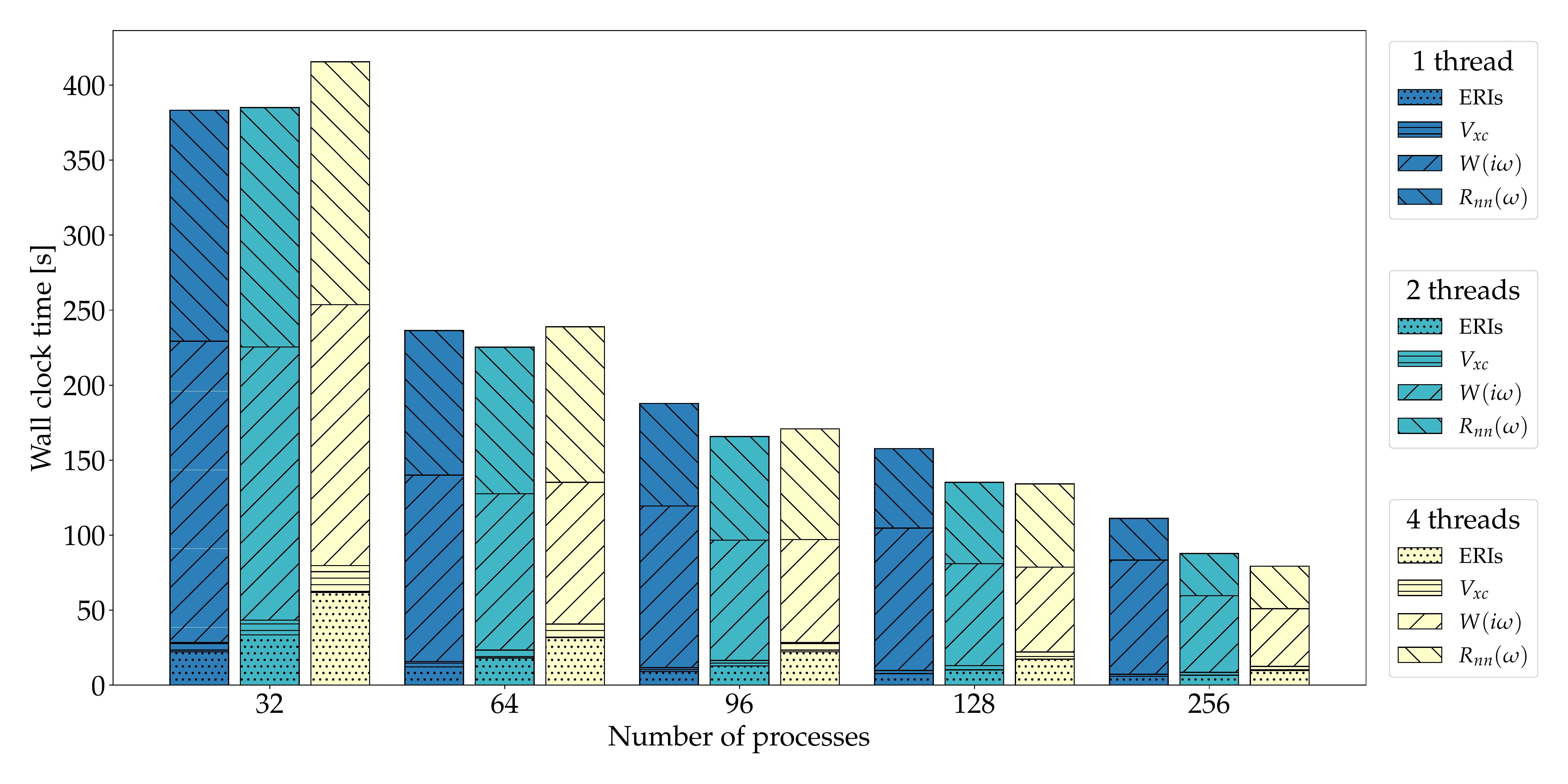}
    \caption{Wall clock time, in seconds, needed to obtain the 5 highest occupied states of the \ce{(H2O)20} cluster with the $G_0W_0$@PBE method.}
    \label{fig:parallel_valence}
\end{figure}

The stacked bars in Figure \ref{fig:parallel_valence} show 
the wall clock time, in seconds, needed to obtain the
$G_0W_0$@PBE energies of the 5 highest occupied states
of \ce{(H2O)20}. Each bar is divided in four sections in order to
reflect the time spent in the most demanding parts of the
CD-$G_0W_0$ implementation. These tasks are 1) ERIs
computation, 2) the exchange-correlation matrix elements,
3) the screened-Coulomb matrix elements on the imaginary
grid $W(i\omega)$, and 4) the computation of the residue
term $R_{nn}(\omega)$. Figure \ref{fig:parallel_valence} also 
shows three categories per processor count, that depend on the 
number of OpenMP threads used per MPI rank. There are several
things to mention about Figure \ref{fig:parallel_valence}. 
First, the ERIs and $V_{xc}$ tasks make
little use of OpenMP parallelization, thus the their
relative times become larger as the number of OpenMP threads
increases. In contrast, both $W(i\omega)$ and $R_{nn}(\omega)$
benefit from the OpenMP parallelization. The overall result is,
in general, a lower execution time for the hybrid OpenMP+MPI approach.
It is also worth noting that both $W(i\omega)$ and $R_{nn}(\omega)$ tasks
have roughly the same computational demand when a few valence
energies are sought. This is the expected behavior
since both tasks have the same $\mathcal{O}(N^4)$ 
asymptotic scaling for valence states. However, the 
computation of core energies dramatically shift the computational burden
to $R_{nn}(\omega)$, as this task has an
$\mathcal{O}(N^5)$ scaling for a core level.
As a result, the overall scaling of CD-$G_0W_0$
can be as high as $N_{core}N_{vir}N_{occ}^2N_{aux}^2$ 
when all $N_{core}$ energies are computed.

The use of the MINRES solver changes the worst scaling to
$N_{core}N_{vir}N_{occ}^2N_{aux}$. MINRES is therefore
an important tool that contributes to the efficiency of the code.
Unfortunately, MINRES can sometimes converge rather slowly. For such cases, the 
Bunch-Kaufman $LDL^T$ factorization might be more efficient. 
Our benchmarks showed that O $2s$ states are problematic for
MINRES, and take most of the total $G_0W_0$ time. Switching to
the $LDL^T$ factorization mildly improves the situation, but the
observed scaling still comes in between $N^5$ and
$N^6$ when the full spectrum is computed.

\begin{figure}
    \centering
    \includegraphics[width=0.45\textwidth]{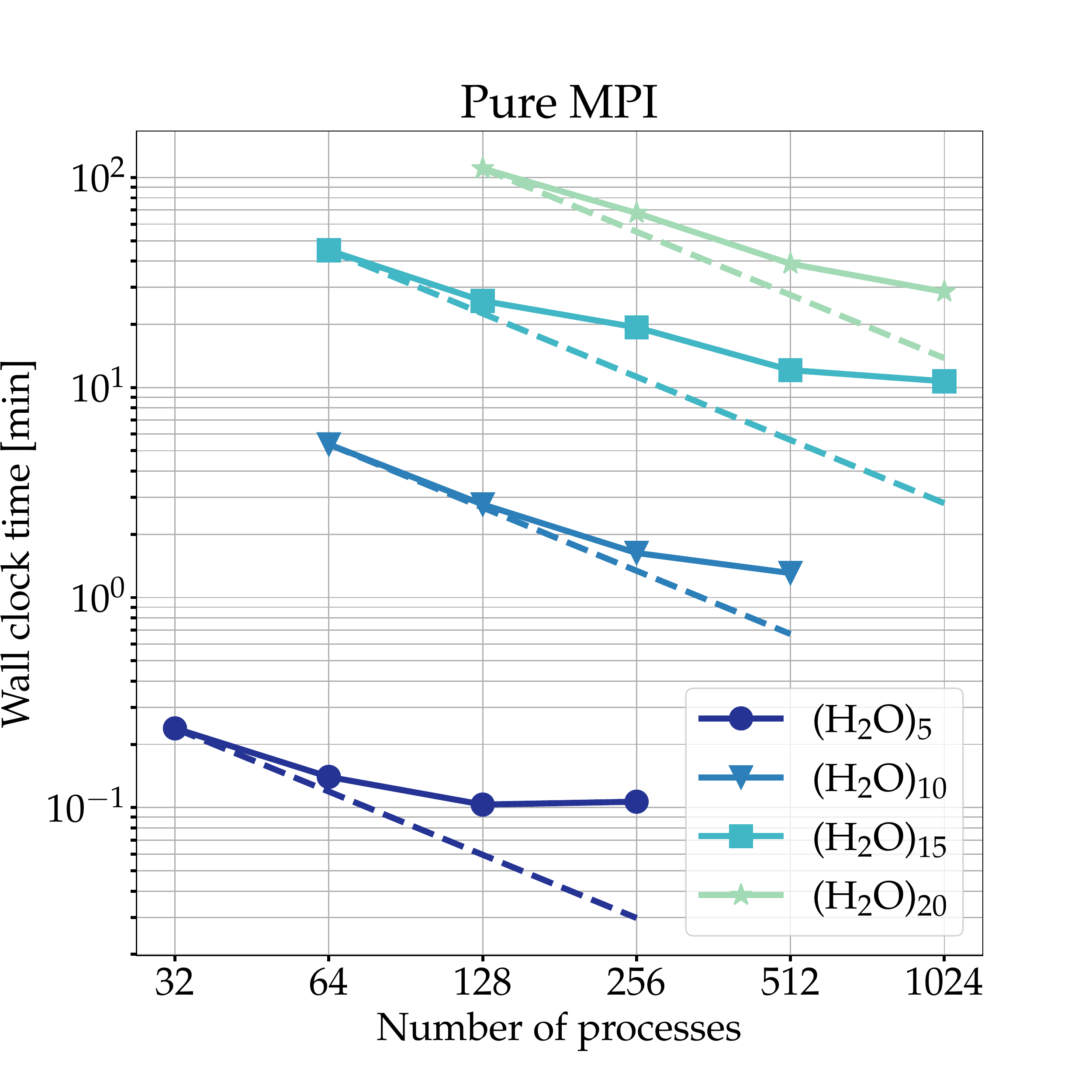}
    \includegraphics[width=0.45\textwidth]{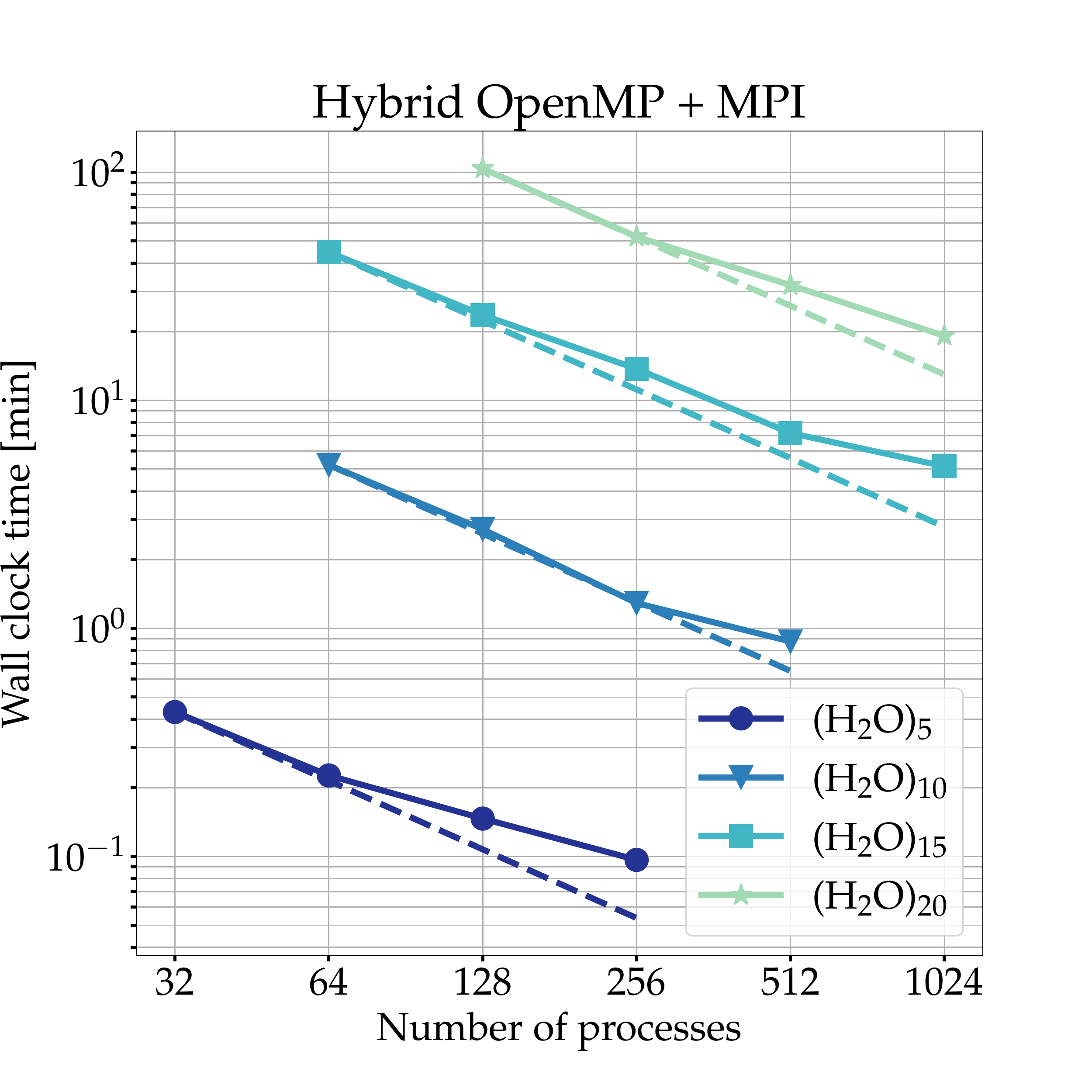}
    \caption{Scaling test with respect to the total number of processes using $G_0W_0$@PBEh to obtain the full occupied spectrum of \ce{(H2O)20}. Four OpenMP threads were used for the hybrid approach and the dashed lines indicate ideal parallelization with respect to the 32-core case.}
    \label{fig:parallel_core}
\end{figure}

This scaling can be inferred from Figure \ref{fig:parallel_core},
which shows the wall clock times needed to obtain all the occupied spectrum
of the water clusters. The execution time increases about 38 times going from \ce{(H2O)10} to \ce{(H2O)20}, which implies an $N^{5.25}$ scaling.
Figure \ref{fig:parallel_core} also shows that the scalability of the
code is greatly enhanced by using the hybrid OpenMP+MPI parallelization.

As noted previously, O $2s$ states mostly determine 
the overall efficiency of the code in this benchmark. 
The use of the $LDL^T$ factorization carries with it both a computation 
and a communication penalty as compared to MINRES. We do not
have a parallel $LDL^T$ factorization at hand to resolve this bottleneck. Still, substituting it with a parallel $LU$ factorization does not show a
computational advantage for these system sizes. 

Several optimized BLAS libraries do have
threaded versions of the $LDL^T$ factorization. As a consequence,
the hybrid OpenMP+MPI is generally the method of choice since it
reduces both the internode communication and the
time spent factorizing $\left[ \mathbf{1} - \bm{\Pi}(\epsilon_m -\omega)\right]^{-1}$.

The performance of the spectral decomposition implementation hinges 
on the performance of the diagonalizer used. We have linked our code to the very efficient ELPA library \cite{ELPA1,ELPA2}
version 2020.11.001 and used the 2-stage solver with the
AVX-512 kernel \cite{ELPA3}. The CD-$GW$ implementation turned out to be
always faster than the spectral decomposition when a few valence energies
are requested. On the other hand, calculations requiring the full
occupied spectrum can be faster with the spectral decomposition
method up to a certain system size. The crossing point occurs at around
15 water molecules when PBE is used as the starting point, and at around 10
water molecules for PBEh. An additional disadvantage
of the spectral decomposition implementation, besides its unfavorable
$\mathcal{O}(N^6)$ scaling, is that the memory scales as $2N_{occ}^2N_{vir}^2$.
More often than not, the lack of memory impedes the use of the
spectral decomposition method.

\section{Summary}
\label{summ}

We have presented and validated a scalable and efficient all-electron $GW$ implementation using Gaussian atomic orbitals to study the valence and core ionization spectroscopies of molecular systems. Our implementation is based on the software infrastructure of the open-source \textsc{NWChem} quantum chemistry package. Specifically, we have implemented the spectral decomposition and contour deformation approaches using the robust variational fitting (RVF) approximation to the four-center electron repulsion integrals. The spectral decomposition approach is linked to the ELPA library for fast eigendecomposition of the Casida matrix, while the contour deformation approach uses the MINRES solver to obtain the action of the inverse dielectric matrices on any given vector. The former approach allows the fast computation of core-level spectroscopy in small molecules with up to 40 atoms, while the latter allows the computation of ionization spectra of molecules with a few hundreds of atoms in reasonable amounts of time.

To validate the accuracy of our implementation, extensive benchmark tests using the 
GW100 and CORE65 datasets and computations of the carbon 1{\it s} binding energy of the well-studied ethyl trifluoroacetate or ESCA molecule were performed. For the ESCA molecule, we compared the $GW$ predictions of the carbon 1{\it s} binding energy using four density functional approximations with experiment.

As a next step of the software development process, we plan to extract our $GW$ implementation to form a
stand-alone domain specific library so that it can be interfaced with any Gaussian basis set based molecular DFT code.
The interface will require as input from the DFT code 
the molecular orbitals and eigenvalue; this can be accomplished by using a variety of data formats
(for example, the Molden format\cite{Schaftenaar2017}).

Extensions to compute neutral excitation spectra based on the BSE formalism are also under development within this framework.

\begin{acknowledgement}

The authors acknowledge funding from the Center for Scalable and
Predictive methods for Excitation and Correlated phenomena (SPEC),
which is funded by the U.S. Department of Energy (DOE), Office of
Science, Office of Basic Energy Sciences, the Division of Chemical
Sciences, Geosciences, and Biosciences. This research also benefited
from computational resources provided by EMSL, a DOE Office of Science
User Facility sponsored by the Office of Biological and Environmental
Research and located at the Pacific Northwest National Laboratory
(PNNL). PNNL is operated by Battelle Memorial Institute for the United
States Department of Energy under DOE contract number
DE-AC05-76RL1830. This research used resources of the National Energy
Research Scientific Computing Center (NERSC), a U.S. Department of
Energy Office of Science User Facility located at Lawrence Berkeley
National Laboratory, operated under Contract No. DE-AC02-05CH11231.

\end{acknowledgement}

\begin{suppinfo}

\begin{itemize}
    \item CORE65 benchmark results at the nonrelativistic ev$GW_0$@PBE,
$G_0W_0$@PBEh, $G_0W_0$@PBE0, and ev$GW_0$@PBE0 levels.
    \item GW100 results for vertical ionization potentials and electron affinities
at the $G_0W_0$@PBE level.
    \item Comparison of core-level binding energies for ethyl-trifluoroacetate using $\Delta$-HF, $\Delta$-PBEh, and
$G_0W_0$@PBEh.
\end{itemize}

\end{suppinfo}

\bibliography{main.bib}

\end{document}